\documentclass[preprint2,twocolumn,times,tighten]{aastex631}
\usepackage{graphicx}	
\usepackage{amsmath}	
\usepackage{amssymb}	

\begin{document}

\title{2D Magnetic Reconnection and Plasmoid Evolution in Partially Ionized Plasmas}

\email{yuehu@ias.edu; *NASA Hubble Fellow}

\author[0000-0002-8455-0805]{Yue Hu*}
\affiliation{Institute for Advanced Study, 1 Einstein Drive, Princeton, NJ 08540, USA}
\affiliation{Cahill Center for Astronomy and Astrophysics, California Institute of Technology, Pasadena, CA 91125, USA}

\author[0000-0002-0458-7828]{Siyao Xu}
\affiliation{Department of Physics, University of Florida, 2001 Museum Rd., Gainesville, FL 32611, USA}

\author[0000-0002-0176-9909]{Grzegorz Kowal}
\affiliation{Escola de Artes, Ci\^encias e Humanidades, Universidade de S\~ao Paulo, Rua Arlindo Bettio 1000, 03828-000, S\~ao Paulo, Brazil}

\author{James M. Stone}
\affiliation{Institute for Advanced Study, 1 Einstein Drive, Princeton, NJ 08540, USA }

\author[0000-0002-7336-6674]{Alex Lazarian}
\affiliation{Department of Astronomy, University of Wisconsin-Madison, Madison, WI, 53706, USA}

\author[0000-0003-3556-6568]{Hui Li}
\affiliation{Theoretical Division, Los Alamos National Laboratory, USA}



\begin{abstract}
Magnetic reconnection in partially ionized plasmas is an important channel for energy release. While the plasmoid instability is well characterized in 2D fully ionized plasmas, its behavior in the presence of neutral-dominated plasma remains poorly understood in the nonlinear, high-Lundquist-number ($S = 10^5$) regime. We present high-resolution ($16384 \times 4096$ cells) isothermal two-dimensional two-fluid (ion $+$ neutral) simulations of Harris-sheet reconnection without ionization or recombination, comparing fully ionized and partially ionized (ionization fraction $\xi = 10^{-1}$ and $10^{-2}$) regimes. Ambipolar drift promotes early current-sheet thinning, while ion-neutral decoupling modifies the subsequent plasmoid evolution. During the simulated interval, the $\xi=10^{-2}$ run remains fragmented into a chain of small plasmoids, whereas the fully ionized run develops a macroscopic ``monster'' plasmoid. Below the neutral-ion decoupling scale $\ell_{\rm dec}$, ions concentrate into the plasmoids, reaching peak overdensities $\rho_i/\rho_{i,0} \approx 10$ ($\xi = 10^{-1}$) and $3-5\times10^{3}$ ($\xi = 10^{-2}$), while the neutrals remain comparatively smooth. This local pile-up raises the ionization fraction and recouples the two fluids within the plasmoids. Measured from the out-of-plane electric field at the reconnection sites, the reconnection rate in the $\xi = 10^{-2}$ case achieves $R_{\rm rec}\approx0.01$, whereas the $\xi = 10^{-1}$ case rises to a rate $\approx0.02$ and further to $0.035$ when apparent coalescence occurs. In the $\xi = 10^{-2}$ case, the ambipolar drift drives a rapid ion inflow $\sim0.5\,v_{A,0}$ into the layer at the same reconnection sites, far above the neutral inflow velocity $\sim0.1\,v_{A,0}$. Here, $v_{A,0}$ is the upstream total Alfv\'en speed. 
\end{abstract}

\keywords{Magnetic fields (994) --- Astrophysical magnetism (102) ---  Plasma astrophysics (1261) --- Magnetohydrodynamics (1964)}

\section{Introduction} \label{sec:intro}

Magnetic reconnection is a fundamental plasma process that converts stored magnetic energy into kinetic and thermal energy through the topological rearrangement of magnetic field lines \citep[see reviews by][]{Biskamp2000, Zweibel_Yamada2009, Yamada2010, 2020PhPl...27a2305L, 2025ARA&A..63..127S}. 
It powers magnetic energy release across the universe---solar flares and coronal mass ejections \citep{Masuda1994, Shibata2011}, gamma-ray burst jets \citep{Drenkhahn_Spruit2002, Lyubarsky2005,2011ApJ...726...90Z,2019ApJ...882..184L}, pulsar magnetospheres \citep{Lyubarsky2001, Cerutti2012}, and the turbulent reconnection diffusion of magnetic 
fields in the interstellar medium and accretion disks \citep{Lazarian_Vishniac1999, Balbus_Hawley1998}. The central theoretical challenge faced by reconnection in astrophysical environments is that: the classical Sweet--Parker model \citep{Sweet1958, Parker1957} predicts a reconnection speed which is smaller than Alfven velocity $V_A$ by a factor $\sim S^{-1/2}$ that, at the astrophysically relevant conditions (Lundquist number $S \gtrsim 10^{10}$), is orders of magnitude too slow to account for the observed dynamical timescales.

In fully ionized plasmas, two distinct mechanisms have been proposed to resolve the slow Sweet--Parker problem: the plasmoid instability and turbulent reconnection. In the plasmoid picture, Sweet--Parker current sheets become unstable to tearing above a critical $S_c \sim 10^4$, fragmenting into a hierarchy of magnetic plasmoids and producing a reconnection rate $V_{\rm rec}/v_{A,0} \sim 0.01$ that is essentially independent of $S$ \citep{Loureiro2007, Bhattacharjee2009, Samtaney2009, Uzdensky2010, Loureiro2012, Comisso2016}. 2D simulations show that the nonlinear state is dominated by plasmoid coalescence and ejection across a wide range of scales, punctuated by occasional macroscopic ``monster'' plasmoids that briefly dominate the energy budget \citep{Huang_Bhattacharjee2010, Huang_Bhattacharjee2013}. In the 3D turbulent picture \citep{Lazarian_Vishniac1999}, pre-existing or self-generated MHD turbulence broadens the effective reconnection layer through stochastic field-line wandering, yielding a fast, resistivity-independent rate \citep{2009ApJ...700...63K}. Nevertheless, recent ultra-high-resolution ($65{,}536^2$ cells) MHD simulations by \citet{2026ApJ..1001..209V} found that the plasmoid-dominated regime is associated with both high Lundquist number and high Reynolds number ($\mathrm{Re} \gtrsim 2000$). Consequently, the reconnection outflows and the fluctuations they drive within and between plasmoids are turbulent \citep{2017ApJ...838...91K},  even in the collisionless regime \citep{2025arXiv251212516H}. The effects of turbulence are essential in 3D reconnection, while we deal in this paper with 2D reconnection. 

Most studies of magnetic reconnection deal with fully ionized plasmas, while many astrophysical environments are not fully ionized. The solar chromosphere \citep{Leake_Arber2006, Arber2007, Ni2015}, dense molecular clouds \citep{1992pavi.book.....S,2008ApJ...677.1151L,2010ApJ...720.1612M,2012ApJ...757..154L}, the warm and cold neutral phases of the ISM \citep{Draine_McKee1993, Ferriere2001}, and protoplanetary disk atmospheres \citep{2018SSRv..214...58B} have ionization fractions $\xi$ ranging from $10^{-7}$ to $10^{-1}$. The presence of neutral-dominated fluid introduces new behaviors: momentum coupling via collisional drag produces ambipolar diffusion of magnetic field and damping of turbulent magnetic fluctuations below the neutral--ion decoupling (i.e., neutral decouples from ion) length $\ell_{\rm dec}$, while above $\ell_{\rm dec}$ the two fluids behave as a single coupled medium \citep{Kulsrud_Pearce1969, Mouschovias1996,1992pavi.book.....S, 2015ApJ...805..118B,2015ApJ...810...44X, 2016ApJ...826..166X, 2024MNRAS.527.3945H}. Whether and how the decoupling effect affects the nonlinear plasmoid regime in 2D is not well understood. 

Earlier studies suggest ambipolar diffusion (AD) can modify the sheet before the tearing instability through the relative ion--neutral drift  \citep{Zweibel1989,VL1999,Leake2012}. Neutral escape via ambipolar drift relaxes the Sweet--Parker mass-conservation constraint, with the asymptotic reconnection rate ultimately set by recombination \citep{VL1999}. In this paper, we do not discuss the effects of recombination, which, similar to turbulence \citep{Lazarian_Vishniac1999, 2004ApJ...603..180L}, increases the reconnection rate in partially ionized gas.

Numerical investigation in 2D has collectively established that ion-neutral collisions reshape plasmoid structure \citep{Leake2012, Ni2015}, the ambipolar drift causes steepening of the field reversal and thinning of the current sheet that enhance plasmoid generation \citep{Ni2015, Murphy2015}, and that ionization-temperature feedback influences the nonlinear evolution \citep{Ni2018_piRecon, Hillier2016}. \citet{2024ApJ...967..136T} developed an analytical theory of tearing onset for a current sheet in partially ionized plasma. More recently, \citet{2026arXiv260223683W} performed a systematic parameter study of ion--neutral collisionality and ionization fraction using a three-fluid five-moment model.

These studies have characterized the linear onset of tearing \citep{2024ApJ...967..136T} and the onset of Hall-mediated fast reconnection at the ion inertial scale \citep{2026arXiv260223683W}, but the fully developed, nonlinear plasmoid stage in partially ionized media remains largely unexplored. In 2D fully ionized plasmas, this nonlinear stage exhibits a set of properties widely regarded as universal: a reconnection rate $V_{\rm rec}/v_{A,0} \sim 0.01$ that is nearly independent of $S$ and the recurrent emergence of ``monster'' plasmoids that dominate the instantaneous energy release \citep{Uzdensky2010, Loureiro2012}. Whether these universal properties survive in neutral-dominated plasma---which decouples from the ion--magnetic field system below the scale $\ell_{\rm dec}$ and damps the flows that drive plasmoid coalescence---is not known, particularly at the high $\beta$ and high Lundquist number. We address this question through 2D high-resolution two-fluid MHD simulations that reach the fully developed plasmoid regime while resolving the neutral-ion decoupling scales. Different from the kinetic, Hall-mediated framework of \citet{2026arXiv260223683W}, our model contains neither the Hall term nor a kinetic inertial scale, so the relevant current-sheet thickness is set instead by ambipolar diffusion and, ultimately, by Ohmic resistivity. 

This paper is organized as follows. In Section~\ref{sec:methods}, we describe the numerical methods, the two-fluid MHD equations, and the initial setup of the Harris current sheet. In Section~\ref{sec:theory}, we discuss the theoretical expectation of reconnection in a partially ionized medium. In Section~\ref{sec:results}, we present our simulation results, highlighting the differences in plasmoid hierarchy and reconnection rate between the fully and partially ionized cases. We discuss the astrophysical implications and summarize our main findings in Section~\ref{sec:discussion} and Section~\ref{sec:conclusion}, respectively.

\section{Numerical Methods} \label{sec:methods}
\subsection{Numerical Implementation} \label{sec:numerics}

We model the partially ionized plasma as two interpenetrating fluids---an ionized component (subscript ``$i$'') and a neutral component (subscript ``$n$'')---coupled through collisional momentum exchange. The magnetic field evolves via the induction equation with explicit Ohmic resistivity. Under the isothermal approximation, the governing two-fluid MHD equations are:
\begin{equation}
\begin{aligned}
  &\frac{\partial \rho_i}{\partial t} 
    + \nabla \cdot (\rho_i \boldsymbol{v}_i) = 0, \\
  &\frac{\partial \rho_n}{\partial t} 
    + \nabla \cdot (\rho_n \boldsymbol{v}_n) = 0,  \\
  &\frac{\partial (\rho_i \boldsymbol{v}_i)}{\partial t} 
    + \nabla \cdot \!\left[ \rho_i \boldsymbol{v}_i \boldsymbol{v}_i 
    + \!\left(c_s^2 \rho_i + \frac{B^2}{8\pi}\right)\! \mathbf{I} 
    - \frac{\mathbf{B}\mathbf{B}}{4\pi} - \mathbf{\Pi}_i \right] \\ 
    & = \gamma_{\rm d} \rho_n \rho_i (\boldsymbol{v}_n - \boldsymbol{v}_i), \\
  &\frac{\partial (\rho_n \boldsymbol{v}_n)}{\partial t} 
    + \nabla \cdot \!\left[ \rho_n \boldsymbol{v}_n \boldsymbol{v}_n 
    + c_s^2 \rho_n \mathbf{I} - \mathbf{\Pi}_n \right] 
    = \gamma_{\rm d} \rho_n \rho_i (\boldsymbol{v}_i - \boldsymbol{v}_n), \\
  &\frac{\partial \mathbf{B}}{\partial t} 
    - \nabla \times (\boldsymbol{v}_i \times \mathbf{B}) 
    = -\nabla \times (\eta \nabla \times \mathbf{B}), \\
  &\nabla \cdot \mathbf{B} = 0,
\end{aligned}
\end{equation}
where $\rho$ and $\boldsymbol{v}$ denote the mass density and velocity of each fluid species, $\mathbf{B}$ is the magnetic field, $c_s$ is the isothermal sound speed, and $\mathbf{I}$ is the identity tensor. $\gamma_{\rm d}$ is the drag coefficient that parameterizes the rate of momentum exchange between the two species by collisional drag force \citep{Draine1986, Draine_McKee1993}. The drag terms are equal and opposite, ensuring total momentum conservation. We adopt an explicit Ohmic resistivity $\eta = 1.0 \times 10^{-5}$ and an isotropic kinematic viscosity $\nu = 1.0 \times 10^{-5}$, applied symmetrically to both fluids through their respective viscous stress tensors $\mathbf{\Pi}_i$ and $\mathbf{\Pi}_n$. This choice yields a magnetic Prandtl number $\mathrm{Pm} \equiv \nu/\eta = 1$. We note that ionization and recombination processes are not included in this study; the ionization fraction is thus set by initial conditions and evolves only through dynamics. 

All simulations are performed using \texttt{AthenaK} \citep{Stone2024_AthenaK}, a performance-portable MHD code. The two-fluid equations are evolved using a third-order implicit--explicit (IMEX3) time integration scheme \citep{Pareschi_Russo2005}, which treats the stiff collisional drag terms implicitly while advancing the hyperbolic fluxes explicitly. This approach is essential for efficiently handling the wide range of coupling timescales encountered across the simulation domain. The Courant--Friedrichs--Lewy (CFL) number is set to 0.3. 

Spatial reconstruction employs the fifth-order WENOZ method \citep{Borges2008}, providing high-order accuracy in smooth regions while capturing shocks and current sheet discontinuities without spurious oscillations. The numerical fluxes are computed using the Local Lax--Friedrichs (LLF) Riemann solver. To ensure numerical stability in the presence of sharp gradients---particularly at the interfaces of plasmoids and within thin current layers---a first-order flux correction is applied locally where needed. The divergence-free constraint on the magnetic field is maintained to machine precision using the constrained transport (CT) algorithm \citep{Evans_Hawley1988, Stone_Gardiner2009}. This is critical for the reconnection problem, where spurious magnetic monopoles can artificially affect the topology of field lines and corrupt the plasmoid dynamics.

\subsubsection{Initial and Boundary Conditions}
We perform two-dimensional simulations of a single Harris current sheet \citep{Harris1962} in a rectangular domain of size $L_x \times L_y = [-1.0, 1.0] \times [-0.25, 0.25]$, resolved by $16384 \times 4096$ computational cells without any refinement. The reconnecting magnetic field profile is given by:
\begin{equation}
  B_x(y) = B_0 \tanh\!\left(\frac{y}{a_0}\right),
  \label{eq:harris}
\end{equation}
where $B_0/\sqrt{4\pi} = 1.0$ is the upstream field strength and $a_0 = 0.01$ is the initial current sheet half-width, corresponding to approximately 80 grid cells. No guide field is applied ($B_z = 0$). The density is initialized with a localized enhancement at the current sheet center with a peak amplitude of 0.5, such that the plasma beta in the upstream inflow region is $\beta\equiv 8\pi (\rho_i+\rho_n) c_s^2 / B_0^2 = 2$, adopting a constant isothermal sound speed of $c_s = 1$. We consider three cases with uniform ionization fraction $\xi = \rho_i/(\rho_i+\rho_n) = 1$, $10^{-1}$, and $10^{-2}$. 

To initiate reconnection, we apply a divergence-free magnetic perturbation $\delta\mathbf{B}$ with amplitude $\epsilon_b = 10^{-4}B_0$ and wavenumber $k_x \approx 20\pi$. In addition, a Gaussian perturbation with amplitude $10^{-6}B_0$ is seeded in the magnetic field to break symmetry and facilitate the growth of secondary tearing instabilities. The imposed narrow-band mode may dominate the early signal and need not be the fastest-growing mode of the evolving sheet.

We adopt periodic boundary conditions along the $x$-direction and damped-flow boundary conditions along the $y$-direction. In the damped-flow boundaries, all variables are smoothly relaxed toward their initial asymptotic upstream values, except for the velocity component $v_y$, which is allowed to pass through freely.

\begin{deluxetable*}{lccccccccccc}
\tablecaption{Summary of simulation parameters and characteristic scales. 
\label{tab:params}}
\tablewidth{0pt}
\tablehead{
  \colhead{Run} & 
  \colhead{$\xi$} & 
  \colhead{$S$} & 
  \colhead{$S_i$} & 
  \colhead{$S_{\rm AD}$} & 
  \colhead{$\delta_{{\rm SP},i}$} & 
  \colhead{$\ell_{{\rm dec},i}$} &
  \colhead{$\delta_{\rm AD}$} & 
  \colhead{$\ell_{\rm dec}$} & 
  \colhead{$a_0$} &
  \colhead{$R_{\rm AD}$} &
  \colhead{Regime}
}
\startdata
$\xi = 1$ & 1.0 & $10^5$ & $10^5$ & --- & 26 & --- & --- & --- & 82 & 0 & Ohmic-dominated \\
$\xi = 10^{-1}$  & 0.1  & $10^5$ & $3.2\times10^5$ & 47 & 15 & 1.8 & 12 & 5 & 82 & 55 & Transitional \\
$\xi = 10^{-2}$ & 0.01 & $10^5$ & $10^6$ & 16 & 8 & 5 & 21 & 50 & 82 & 610 & AD-dominated \\
\enddata
\tablecomments{All length scales are given in units of computational cells. $S_i = Lv_{A,i}/\eta$ is the ion Lundquist number; it governs tearing only in the dynamically decoupled limit. Here $v_{A,i}$ is the ion Alfv\'en speed. $S_{\rm AD} = a_0 v_{A,i,0}/\eta_{\rm AD}$ is the ambipolar Lundquist number. $\delta_{{\rm SP},i} = L/\sqrt{S_i}$ is the ion-only Sweet--Parker thickness. $\ell_{{\rm dec},i}$ and $\ell_{\rm dec}$ are the ion and neutral decoupling scales (Eqs.~\ref{eq:ldion} and~\ref{eq:ldn}). $\delta_{\rm AD} = a_0/\sqrt{S_{\rm AD}}$ is a characteristic ambipolar transport scale. Note that $S_{\rm AD}$ and $S$ are defined for different physical processes, and the two are therefore not directly comparable.
}
\end{deluxetable*}

\section{Theoretical consideration}
\label{sec:theory}
Several key lengths and times govern the physics of reconnection in a partially ionized medium. Taking the half-length of the current sheet $L = L_x/2 = 1.0$ as the macroscopic scale and the total Alfv\'{e}n speed $v_{A} \equiv B_0/\sqrt{4\pi(\rho_{i}+\rho_n)}$ as the characteristic velocity, the system is parameterized by the Lundquist number \citep{Sweet1958, Parker1957}:
\begin{equation}
  S \equiv \frac{L \, v_{A}}{\eta} \sim 10^5,
  \label{eq:lundquist}
\end{equation}
which places our simulations close to the critical threshold for the plasmoid instability ($S_c \sim 10^4$; \citealt{Bhattacharjee2009}), ensuring that the current sheet fragments into a fully developed plasmoid chain.

For time normalization, we define $t_{A,0}=L/v_{A,0}$ and $t_{A,i}=L/v_{A,i,0}$ using the initial upstream total-density and ion Alfv\'en speeds, respectively. For the initial uniform ionization fraction $\xi_0=\rho_{i,0}/(\rho_{i,0}+\rho_{n,0})$, these times satisfy $t_{A,i}=\sqrt{\xi_0}\,t_{A,0}$. Thus, $5t_{A,i}$ corresponds to approximately $1.58t_{A,0}$ and $0.50t_{A,0}$ for $\xi_0=10^{-1}$ and $10^{-2}$, respectively. Unless otherwise stated, simulation times are expressed in units of $t_{A,i}$; the modal-growth comparisons use the common normalization $t_{A,0}$.


\subsection{Ion--Neutral Coupling and Characteristic Scales} \label{sec:coupling}
The collisional coupling between the ion and neutral fluids is characterized by two momentum-exchange frequencies \citep{Shu1983}. The ion-neutral collision frequency (the rate at which an ion transfers its momentum to neutrals) is:
\begin{equation}
  \nu_{in} = \gamma_{\rm d} \rho_n,
  \label{eq:nuin}
\end{equation}
and the neutral--ion collision frequency (the rate at which a neutral is dragged by ions) is:
\begin{equation}
  \nu_{ni} = \gamma_{\rm d} \rho_i.
  \label{eq:nuni}
\end{equation}
By momentum conservation, the two frequencies satisfy $\rho_i \nu_{in} = \rho_n \nu_{ni}$. For a weakly ionized medium ($\xi \ll 1$), the relation $\nu_{ni} \ll \nu_{in}$ holds: ions are frequently deflected by the dominant neutral population, while any given neutral atom rarely encounters an ion. Physically, $\gamma_{\rm d}$ is set by the momentum-transfer rate coefficient for ion--neutral elastic scattering \citep{Draine1986}. 

\subsubsection{Ambipolar Diffusion} 
\label{sec:AD}
In the coupled regime, ion--neutral drag permits a relative drift between the ionized and neutral fluids. Its effect on magnetic-flux transport is conventionally characterized by the ambipolar-diffusion coefficient \citep{Mestel_Spitzer1956,Shu1983,Zweibel2002},
\begin{equation}
  \eta_{\rm AD}
  = \frac{B^2}{4\pi\,\gamma_{\rm d}\rho_i\rho_n}
  = \frac{v_{A,i}^2}{\gamma_{\rm d}\rho_n}
  = v_{A,i}\ell_{{\rm dec},i}.
  \label{eq:etaAD}
\end{equation}
Ambipolar drift transports magnetic flux relative to the neutral fluid without directly violating flux freezing in the ion fluid. The associated ion--neutral drag dissipates the relative kinetic energy at a rate
\begin{equation}
Q_{\rm drag}
=
\gamma_{\rm d}\rho_i\rho_n
|\boldsymbol{v}_i-\boldsymbol{v}_n|^2,
\end{equation}
thereby allowing magnetic energy transferred into ion--neutral drift to be converted into heat. However, ambipolar drift by itself does not change magnetic topology; the non-ideal electric field required for reconnection is supported by Ohmic resistivity. The relative strength of ambipolar transport to Ohmic dissipation is parametrized by the ambipolar Reynolds number
\begin{equation}
  R_{\rm AD} \equiv \frac{\eta_{\rm AD}}{\eta} 
  = \frac{v_{A,i}^2}{\gamma_d \rho_n \,\eta},
  \label{eq:R_AD}
\end{equation}
with $R_{\rm AD} \gg 1$ corresponding to the regime where ambipolar drift controls the macroscopic flux transport while Ohmic dissipation, confined to substantially thinner inner layers, mediates the topological reconnection.

A characteristic transverse length scale associated with this transport can be defined as the distance over which ambipolar diffusion acts during an Alfv\'en crossing time across the initial sheet width, $t_{A,a}=a_0/v_{A,i}$. Because this scale characterizes the local modification of the transverse current-sheet profile, the appropriate reference length is $a_0$:
\begin{equation}
  S_{\rm AD} \equiv \frac{a_0 \, v_{A,i}}{\eta_{\rm AD}},
  \qquad
  \delta_{\rm AD} = \frac{a_0}{\sqrt{S_{\rm AD}}}.
  \label{eq:delta_AD}
\end{equation}
We stress that $\delta_{\rm AD}$ is not a dissipation-set sheet width in the Sweet--Parker sense, but a geometric scale characterizing the ambipolar drift-mediated configuration of the current sheet. The actual resistive inner layer where Ohmic dissipation operates lies within $\delta_{\rm AD}$ and is set by $\eta$. The position of $\delta_{\rm AD}$ relative to the two decoupling scales determines whether the ambipolar drift zone overlaps with, contains, or lies outside the resistive layer. 

Throughout this paper, ``AD-mediated'' effects refer to magnetic-flux transport and current-sheet configuration produced by ion--neutral drift. The associated drag dissipates relative kinetic energy at the rate $Q_{\rm drag}$ given above, while changes in magnetic connectivity require the Ohmic non-ideal electric field in this model.

\subsubsection{Onset stage} 
The \emph{neutral decoupling scale}, $\ell_{\rm dec}$, marks the onset of significant two-fluid effects, below which neutrals can no longer remain dynamically coupled to the ion--magnetic-field system. In general, $\ell_{\rm dec}$ is determined by equating the characteristic dynamical frequency of the physical process of interest at scale $\ell$ to the neutral--ion collision frequency,
\begin{equation}
\nu_{ni}=\gamma_{\rm d}\rho_i.
\end{equation}

For magnetic reconnection, the relevant dynamical frequency depends on the physical process controlling each evolutionary stage:
\begin{itemize}
    \item During the pre-tearing onset stage, the question is whether neutrals can follow the magnetically driven thinning of the current sheet, so the relevant frequency is the Alfv\'en-crossing frequency of the coupled fluid.
    \item During linear tearing, coupling is instead determined by whether neutrals can respond over an instability growth time, requiring comparison of $\nu_{ni}$ with the tearing growth rate. 
    \item In the nonlinear stage, the relevant processes are the transport of ions and neutrals into and out of the reconnection layer, so $\nu_{ni}$ should be compared with the local inflow and outflow rates.
\end{itemize}  

Here, we focus on the onset-stage criterion. Approximating the relevant dynamical frequency by the Alfv\'en-crossing rate, the neutral-ion decoupling scale in the pre-tearing stage is
\begin{equation}
\ell_{\rm dec}
=\frac{v_A}{\nu_{ni}}
=\frac{v_A}{\gamma_{\rm d}\rho_i}.
\label{eq:ldn}
\end{equation}
At scales $\ell>\ell_{\rm dec}$, the collision frequency exceeds the Alfv\'en-crossing frequency, and ions and neutrals behave approximately as a single fluid. At $\ell<\ell_{\rm dec}$, neutrals can no longer follow the magnetically driven ion motions. The Alfv\'enic fluctuations then enter a strongly damped regime in which ion motions are impeded by frictional drag against the neutral background \citep{2016ApJ...826..166X,2024MNRAS.527.3945H}.

The \emph{ion decoupling scale} $\ell_{{\rm dec},i}$ marks the lower boundary of this damped zone, where the Alfv\'en-crossing frequency equals the ion--neutral collision frequency $\nu_{in} = \gamma_{\rm d} \rho_n$:
\begin{equation}
  \ell_{{\rm dec},i} = \frac{v_{A,i}}{\nu_{in}} 
  = \frac{v_{A,i}}{\gamma_{\rm d}\rho_n}.
  \label{eq:ldion}
\end{equation}
At scales $\ell < \ell_{{\rm dec},i}$, the ion collisions with neutrals cannot appreciably affect their dynamics: the ions are effectively decoupled from the neutrals and Alfv\'{e}n waves propagate at the fast ion Alfv\'en speed $v_{A,i} = B/\sqrt{4\pi\rho_i}$, with the neutrals acting as a passive, non-participating background. Based on the decoupling scales, we have three distinct dynamic regimes:
\begin{equation}
  \underbrace{\ell > \ell_{\rm dec}}_{\text{coupled: } v_{A}}
  \quad \longrightarrow \quad
  \underbrace{\ell_{{\rm dec},i} < \ell < \ell_{\rm dec}}_{\text{damped (AD zone)}}
  \quad \longrightarrow \quad
  \underbrace{\ell < \ell_{{\rm dec},i}}_{\text{decoupled: } v_{A,i}}.
  \label{eq:three_regimes}
\end{equation}
The intermediate ``damped zone'' between $\ell_{{\rm dec},i}$ and $\ell_{\rm dec}$ is the range of scales where ambipolar drift dominates the modification of the magnetic field configuration, and its width in logarithmic scale space is $\log(\ell_{\rm dec}/\ell_{{\rm dec},i}) = \log[(1-\xi)/\sqrt{\xi}]$. 


The relative ordering of the characteristic scales determines the reconnection regime. We now compute each scale explicitly for our simulations. In our numerical setup (see Table~\ref{tab:params}), the drag coefficient $\gamma_{\rm d}$ is chosen such that $\ell_{\rm dec} \approx 50$ cells and $\ell_{{\rm dec},i} \approx 5$ cells for $\xi = 10^{-2}$. This ensures that the reconnection starts in the fully coupled regime, as the current sheet width is $\approx 80$ cells, and gradually transits to the damped and decoupled regimes. For $\xi = 10^{-1}$, we have $\ell_{\rm dec} \approx 5$ cells and $\ell_{{\rm dec},i} \approx 1.8$ cells such that ion and neutral are dynamically coupled.

\paragraph{Ohmic Sweet--Parker thickness.}
For the fully ionized reference case, the Sweet--Parker current sheet thickness based on the Ohmic resistivity is:
$\delta_{\rm SP} = \frac{L}{\sqrt{S}}\approx 26~{\rm cells}$. The initial current sheet half-width satisfies $a_0 > \delta_{\rm SP}$, guaranteeing that the sheet must thin before steady-state reconnection can proceed. Since $S = 10^5 > S_c$, the plasmoid instability is expected during this thinning process \citep{Bhattacharjee2009, Uzdensky2010}.
For comparisons with the partially ionized runs, we distinguish the total-density reference scale, $\delta_{{\rm SP},0}\equiv L/\sqrt{S}$, from the ion-only reference scale, $\delta_{{\rm SP},i}=L/\sqrt{S_i}$. For $\xi=10^{-2}$, their ordering relative to the ambipolar transport estimate is $\delta_{{\rm SP},i}\simeq8 <\delta_{\rm AD}\simeq21 <\delta_{{\rm SP},0}\simeq26$ cells. The transport estimate $\delta_{\rm AD}$ should be distinguished from the measured current-sheet half-width $a_{\rm eff}$ introduced in Appendix~\ref{sec:tearing_estimates}.

\paragraph{Ambipolar diffusion thickness.}
For the partially ionized cases, 
the position of $\delta_{\rm AD}$ relative to the two decoupling scales $\ell_{{\rm dec},i}$ and $\ell_{\rm dec}$ is the key diagnostic for the reconnection regime:
\begin{itemize}
    \item \textbf{$\xi = 10^{-2}$ (AD-dominated regime):}
    \begin{equation}
      \ell_{{\rm dec},i}(5) \;<\; \delta_{\rm AD}(21) \;<\; \ell_{\rm dec}(50) 
      \;<\; a_0(82)
      \quad [\text{cells}].
      \label{eq:hierarchy_001}
    \end{equation}
    The ambipolar thickness $\delta_{\rm AD}$ falls \emph{between} the two decoupling scales, placing the reconnection inner layer squarely within the ambipolar damping zone. In this regime, the neutrals have already decoupled at the current sheet scale ($\ell_{\rm dec} < a_0$), so even the global structure of the sheet is influenced by two-fluid effects. Although the neutrals do not directly participate in the magnetic field dynamics, they indirectly control the geometry of the reconnection layer through frictional drag.
    \item \textbf{$\xi = 10^{-1}$ (transitional regime):}
    \begin{equation}
      \ell_{{\rm dec},i}(1.8) \;<\; \ell_{\rm dec}(5) \;<\; \delta_{\rm AD}(12) 
      \;<\; a_0(82)
      \quad [\text{cells}].
      \label{eq:hierarchy_01}
    \end{equation}
    Here $\delta_{\rm AD}$ lies \emph{above} both decoupling scales. The reconnection inner layer extends beyond the ambipolar damping zone into the coupled regime, placing the system in a transitional state between AD-dominated and purely resistive reconnection. The narrower damping zone means that the range of scales where AD operates is relatively limited. Nevertheless, $\eta_\text{AD}/\eta \approx 68$ remains large, so AD-mediated flux transport remains a significant contribution to the overall reconnection dynamics. The weaker AD means the transition to the plasmoid-dominated phase is delayed relative to $\xi = 10^{-2}$. This intermediate regime bridges the fully ionized limit---where monster plasmoids freely form and merge---and the strongly decoupled $\xi = 10^{-2}$ limit where the plasmoid hierarchy is truncated at $\ell_{\rm dec}$.
\end{itemize}

\begin{figure*}
\centering
\includegraphics[width=0.99\linewidth]{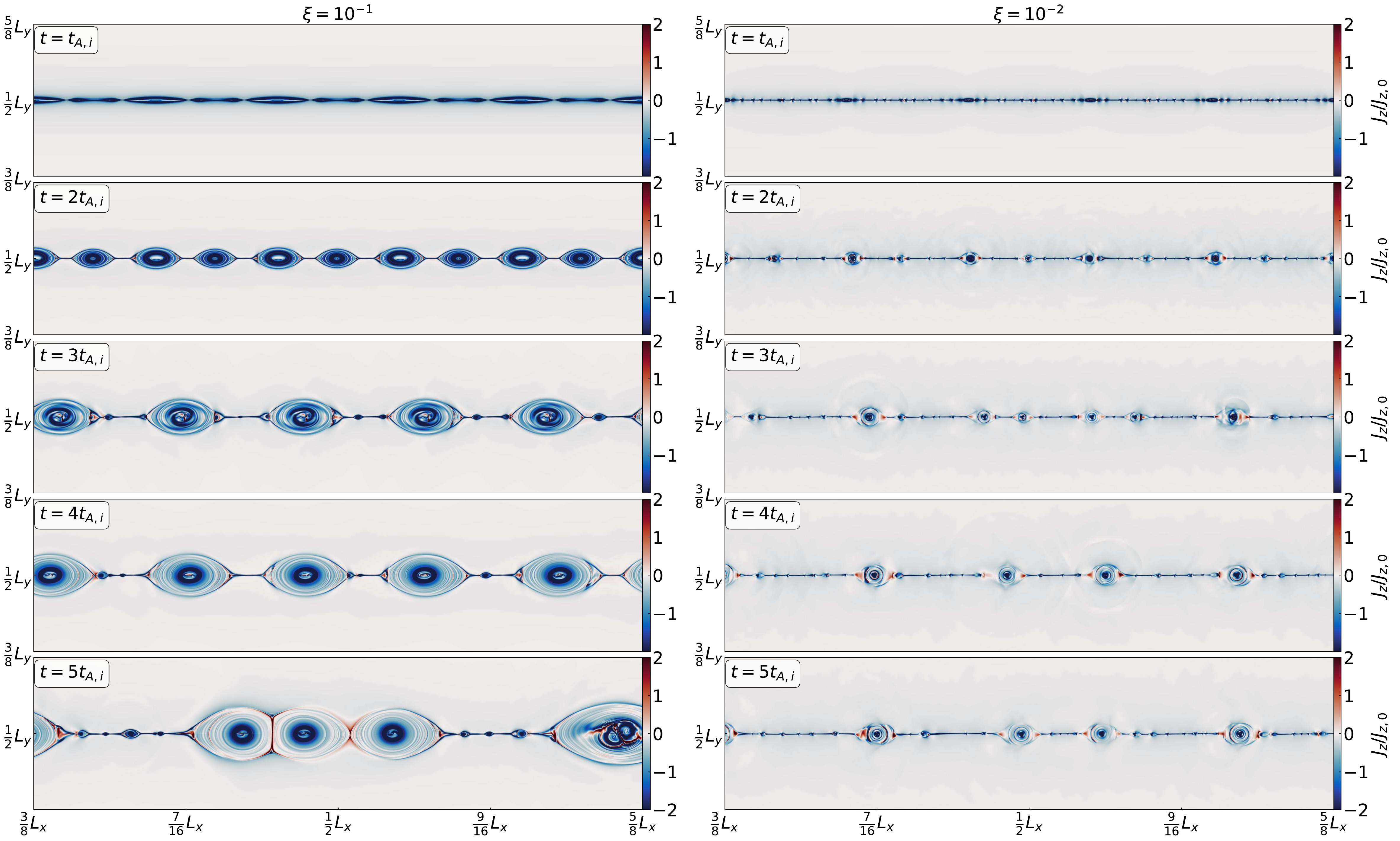}
        \caption{Time evolution of the normalized out-of-plane current density ($J_z/J_{z,0}$) during the magnetic reconnection process. The left and right columns correspond to simulations with initial ionization fractions of $\xi = 10^{-1}$ and $\xi = 10^{-2}$, respectively. Snapshots are taken at consecutive times $t = 1, 2, 3, 4$, and $5 t_{A,i}$ (from top to bottom). Consistent with Fig.~\ref{fig:CS_rho}, only the central $1/4$ of the computational domain is shown to highlight the detailed dynamics within the reconnecting current sheet. Here, $J_{z,0}$ denotes the initial peak current density.}
    \label{fig:CS_Jz}
\end{figure*}

\begin{figure*}
\centering
\includegraphics[width=0.99\linewidth]{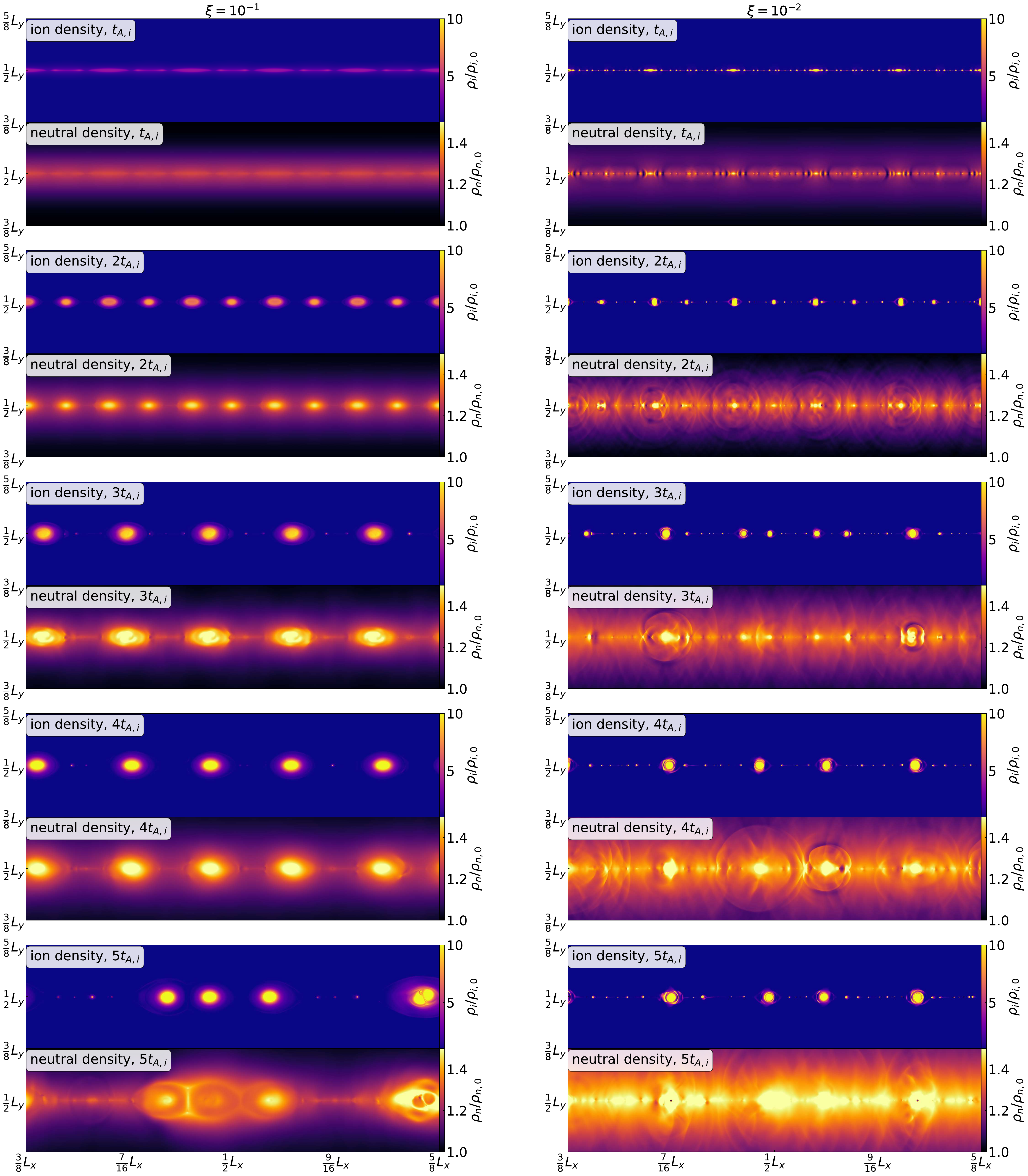}
        \caption{Time evolution of the spatial distributions of the normalized ion density ($\rho_i/\rho_{i,0}$) and neutral density ($\rho_n/\rho_{n,0}$) during the magnetic reconnection process. The left column presents the simulation case with an initial ionization fraction of $\xi = 10^{-1}$, while the right column displays the case with $\xi = 10^{-2}$. From top to bottom, the panels show snapshots at consecutive times: $t = 1, 2, 3, 4$, and $5 t_{A,i}$, where $t_{A,i}$ is the ion Alfv\'en crossing time. To clearly visualize the dynamics within the current sheet, including the formation and merging of plasmoids, only the central $1/4$ of the computational domain is shown. The subscript "0" denotes the initial upstream values.}
    \label{fig:CS_rho}
\end{figure*}

\begin{figure*}
\centering
\includegraphics[width=0.99\linewidth]{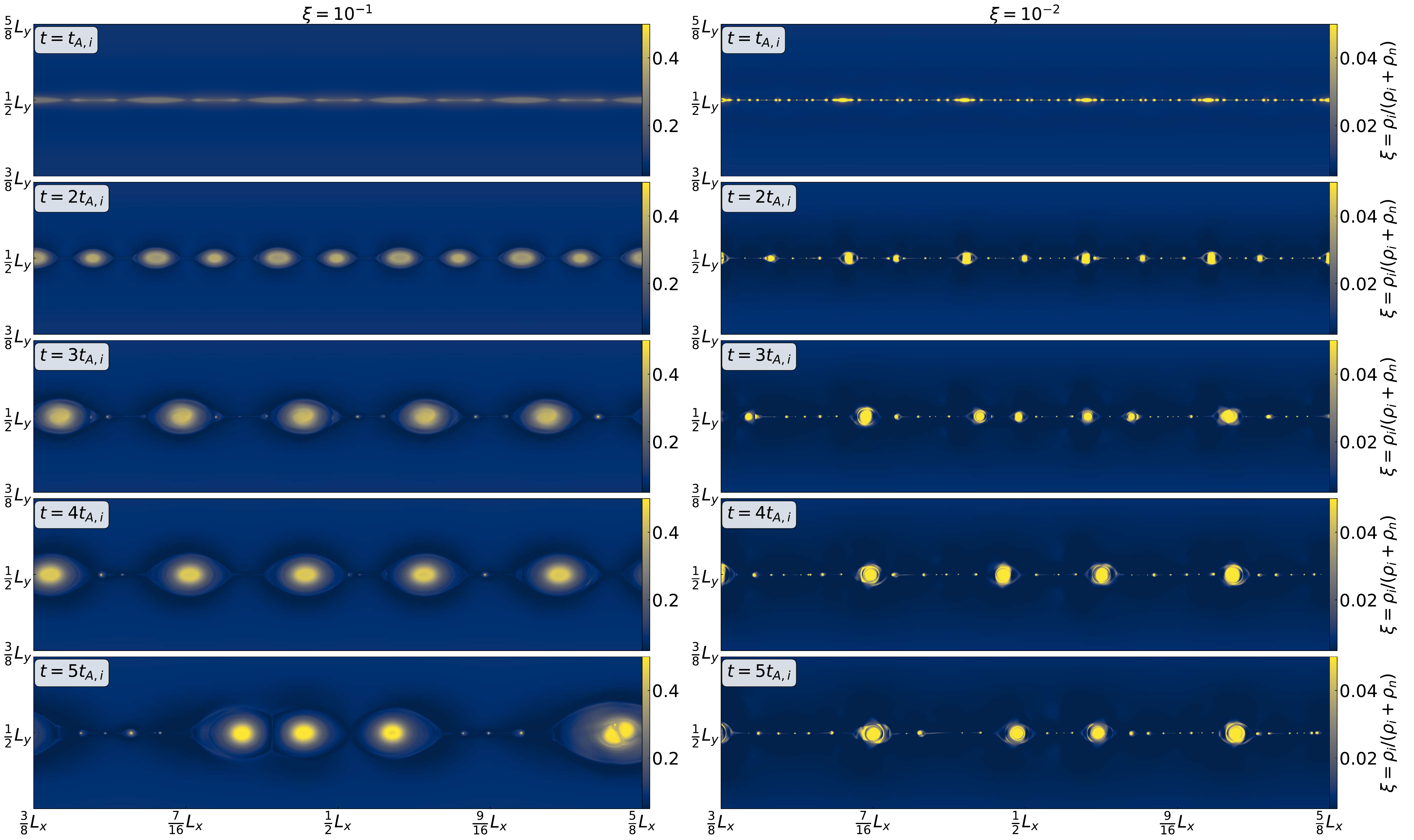}
        \caption{Same as Fig.~\ref{fig:CS_Jz}, but for the ionization fraction.}
    \label{fig:CS_xi}
\end{figure*}

\begin{figure*}
\centering
\includegraphics[width=0.99\linewidth]{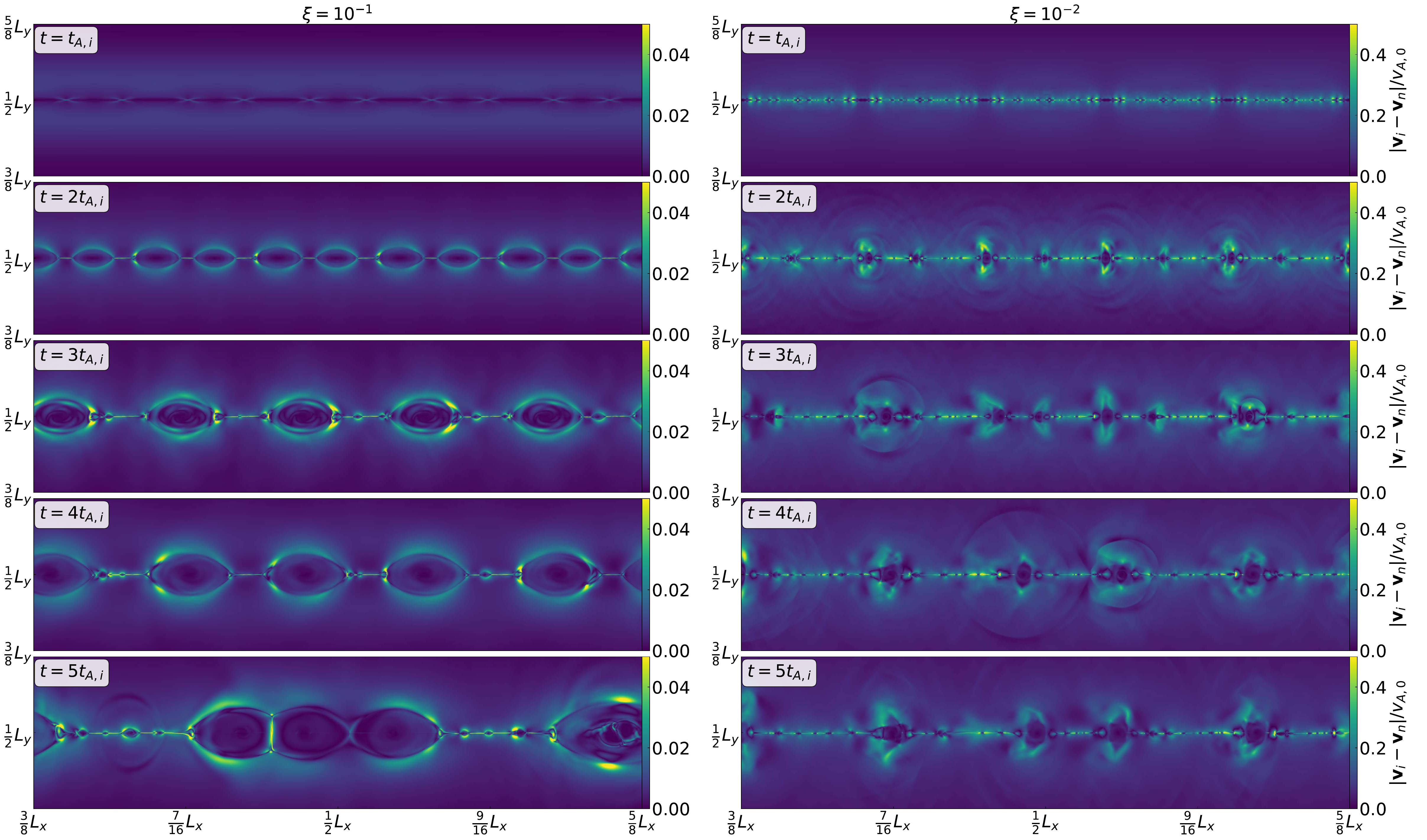}
        \caption{Same as Fig.~\ref{fig:CS_Jz}, but for the normalized drift velocity. $v_{A, 0}$ is  the initial upstream total Alfv\'en speed.}
    \label{fig:CS_drift}
\end{figure*}

\subsubsection{Ion--neutral coupling during linear tearing}
\label{sec:tearing_coupling}

For a tearing mode with growth rate $\gamma$, the ion--neutral response is coupled for $\gamma\ll\nu_{ni}$, intermediate for $\nu_{ni}\ll\gamma\ll\nu_{in}$, and decoupled for $\gamma\gg\nu_{in}$ \citep{Zweibel1989,Pucci2020}. The effective inertia participating in the instability is
\begin{equation}
\rho_{\rm eff}
=
\rho_i\left(1+\frac{\nu_{in}}{\gamma+\nu_{ni}}\right),
\label{eq:tearing_inertia}
\end{equation}
which approaches $\rho_i+\rho_n$ and $\rho_i$ in the coupled and decoupled limits, respectively.

For a quasi-stationary Harris-like sheet of half-width $a$, the fastest-mode growth rate can be written as
\begin{equation}
\gamma_{\max}\sim
\left(\frac{\eta v_{A,i}}{a^3}\right)^{1/2}
\left(1+\frac{\nu_{in}}{\gamma_{\max}+\nu_{ni}}\right)^{-1/4},
\label{eq:tearing_growth_inertia}
\end{equation}
which is obtained by matching the small- and large-$\Delta'$ branches (see Appendix \ref{sec:tearing_estimates}). Here, the Alfv\'en speed and collision frequencies are evaluated using the magnetic field immediately outside the sheet and representative local ion and neutral densities. In the coupled limit,
\begin{equation}
\gamma_{\max}\sim
\left(\frac{\eta v_A}{a^3}\right)^{1/2},
\qquad
k_{\max}a\sim
\left(\frac{av_A}{\eta}\right)^{-1/4},
\label{eq:coupled_tearing_scaling}
\end{equation}
where $v_A$ is based on the total density. Thus, at fixed field strength and density, ambipolar-drift-driven sheet thinning increases the tearing rate as $\gamma_{\max}\propto a^{-3/2}$ until the coupled approximation breaks down. The corresponding coupled ideal-tearing threshold is $a_{\rm IT}/L\sim S^{-1/3}$.

The transition from coupled to intermediate tearing occurs when $\gamma_{\max}\sim\nu_{ni}$, defining
\begin{equation}
a_{\rm tr}\sim
\left(\frac{\eta v_A}{\nu_{ni}^2}\right)^{1/3},
\qquad
\frac{\gamma_{\max}}{\nu_{ni}}
\sim
\left(\frac{a_{\rm tr}}{a}\right)^{3/2}
\quad (a\gg a_{\rm tr}).
\label{eq:tearing_coupling_width}
\end{equation}
Ion accumulation, however, can increase $\nu_{ni}$ and shift the crossover to smaller $a_{\rm tr}$, so the coupling regime at tearing onset depends on the locally evolved, ambipolar-drift-modified sheet rather than on the initial ionization fraction alone. Intermediate coupling can additionally introduce frictional stabilization \citep{Pucci2020}. These estimates use inviscid scalings and omit order-unity coefficients; viscosity at $\mathrm{Pm}=1$ and departures from a Harris profile can modify the quantitative growth rates.

\subsubsection{Nonlinear stage}
In the nonlinear stage, the local advective rate determines neutral-ion decoupling. For inflow across the reconnection layer, the corresponding decoupling scale is
\begin{equation}
\ell_{\rm dec,nl}(\boldsymbol{x},t)
\sim
\frac{|v_{\rm in}(\boldsymbol{x},t)|}
{\gamma_{\rm d}\rho_i(\boldsymbol{x},t)}.
\label{eq:ldec_nl}
\end{equation}
Using the upstream density, a site-averaged maximum inflow speed of $\sim0.1v_A$ gives a characteristic decoupling scale approximately one order of magnitude smaller than the onset-stage estimate. This is only an order-of-magnitude estimate because both the inflow speed and ion density vary spatially and temporally. Once decoupled, the neutrals no longer follow the ion inflow and instead develop their own dynamics. However, the different dynamics can produce strong ion pile-up in the reconnection layer \citep{VL1999}, increasing the local collision frequency and reducing the decoupling scale.


\section{Results}
\label{sec:results}
\subsection{Plasmoid Morphology and Neutral-Ion Decoupling}
Fig.~\ref{fig:CS_Jz} shows the time evolution of the normalized out-of-plane current density $J_z/J_{z,0}$ for the two partially ionized runs, and Fig.~\ref{fig:CS_rho} shows the corresponding normalized ion and neutral mass densities. Figs.~\ref{fig:CS_xi} and \ref{fig:CS_drift} present the spatial distributions of the local ionization fraction $\xi(x,y,t) \equiv \rho_i/(\rho_i + \rho_n)$ and of the normalized ion--neutral drift velocity $|\boldsymbol{v}_i - \boldsymbol{v}_n|/v_{A,0}$, respectively, where $v_{A,0}$ is the initial upstream total Alfv\'en speed. Appendix Fig.~\ref{fig:CS_xi1} shows the fully ionized reference run.

\paragraph{AD-dominated regime ($\xi = 10^{-2}$):}
The most weakly ionized run, in which the characteristic ambipolar transport scale lies between the two Alfv\'en-crossing decoupling scales, develops a morphology qualitatively distinct from the fully ionized baseline (Appendix Fig.~\ref{fig:CS_xi1}). Already at $t = t_{A,i}$ (Fig.~\ref{fig:CS_Jz}, right column), the current sheet is populated by a dense array of small-amplitude plasmoids, indicating that the linear instability saturates within a single ion Alfv\'en crossing time. In the subsequent non-linear phase ($t_{A,i} \le t \le 5\,t_{A,i}$), the relative ion--neutral drift efficiently dissipates the ion motions that drive coalescence, causing the chain to undergo small-scale mergers. Still, it is difficult to develop the hierarchical plasmoid cascade, i.e., smaller plasmoids formed in secondary current sheets, and no significant macroscopic plasmoid emerges over the duration of the run (see Appendix Fig.~\ref{fig:plasmoid_stats} for the details of plasmoid statistics).

Different from ions that are concentrated into narrow current sheets, neutrals are spatially more extended.  Ions accumulate preferentially in plasmoid cores, raising the local ionization fraction and decreasing the decoupling length $\ell_{\rm dec} \propto \rho_i^{-1}$. This local re-coupling is directly visualized in Fig.~\ref{fig:CS_xi} (right column): although the upstream value is $\xi = 10^{-2}$, plasmoid cores reach $\xi \approx 1$ by $t = 3\,t_{A,i}$, a two-orders-of-magnitude enhancement that proportionally shortens the local $\ell_{\rm dec}$ to a scale comparable to or below the plasmoid size. The neutrals are therefore re-coupled to the ions in and around plasmoid cores, even as they remain effectively decoupled in the dilute plasmoid peripheries. The observed neutral morphology in Fig.~\ref{fig:CS_rho} is consequently a composite of two regimes: wake-like structures appear in the neutral distribution around the contracting, moving, and merging plasmoid and ion-coupled dense features at the plasmoid cores themselves.

The drift velocity field (Fig.~\ref{fig:CS_drift}, right column) provides the most direct diagnostic of where decoupling is dynamically significant. Peak values of $|\boldsymbol{v}_i - \boldsymbol{v}_n|/v_{A,0}\approx 0.4$ are reached along the peripheries of contracting plasmoids, where the Lorentz force on the ions is strongest while neutrals lag. Within plasmoid cores, by contrast, $|\boldsymbol{v}_d|$ drops to near zero, consistent with the local re-coupling driven by the elevated $\xi$ in those regions (Fig.~\ref{fig:CS_xi}). We also see that the steep gradients in the drift velocity distribution trace the expanding wave front and trailing wakes in neutrals associated with contracting and moving plasmoids, respectively.

This spatial heterogeneity of the coupling produces a dynamical signature: concentric, ring-like fluctuations emanate from individual plasmoids and propagate outward into the upstream (Fig.~\ref{fig:CS_rho}, right column, $t = 2\text{--}5\,t_{A,i}$). The rings appear only in the neutral density --- they are not seen in the co-spatial ion fluid, nor in the $\xi = 10^{-1}$ neutral panels --- and we identify them as fluctuations driven by plasmoid contractions and mergers. Because the neutrals are locally locked to the dense plasmoid cores, they are advected with the plasmoids and impulsively distributed when two plasmoids coalesce. The resulting neutral response propagates outward into the low-ion-density region, where the neutral-ion coupling is weak. As successive mergers occur, the wavefronts broaden and overlap, and by $t = 5\,t_{A,i}$ the upstream neutral medium has reorganized into a diffuse, structured background of interfering waves.

\paragraph{Transitional regime ($\xi = 10^{-1}$):}
Increasing the ionization fraction by an order of magnitude to $\xi = 10^{-1}$ places the system in the transitional regime, in which the ambipolar inner-layer width exceeds the neutral decoupling scale. We see a broader primary current sheet compared to the case with $\xi = 10^{-2}$. By $t = 2\,t_{A,i}$ the sheet has fragmented into a plasmoid chain of comparable size, separated by thin secondary current sheets clearly visible in Fig.~\ref{fig:CS_Jz} (left column). In contrast to the $\xi = 10^{-2}$ case, coalescence now proceeds vigorously: between $t = 3\,t_{A,i}$ and $t = 5\,t_{A,i}$ the chain relaxes through successive mergers into several well-separated plasmoids of growing size, with a precursor ``monster'' plasmoid emerging in the rightmost portion of the domain at $t = 5\,t_{A,i}$. The peak neutral enhancements saturate at $\rho_n/\rho_{n,0} \approx 1.4$--$1.7$, while the corresponding ion peaks reach $\rho_i/\rho_{i,0} \approx 5$--$12$. This $\sim$10-fold differential is the expected signature of the transitional regime --- the neutrals are coupled to the global flow through the drag term but cannot follow the fine-scale (i.e., smaller than $\ell_{\rm dec}$) density fluctuations.

The ionization fraction within plasmoid cores rises to $\xi \approx 0.4$--$0.5$ (Fig.~\ref{fig:CS_xi}, left column), a four-to-five-times enhancement. With this strong local re-coupling, the drift velocity (Fig.~\ref{fig:CS_drift}, left column) remains modest throughout the domain, with peak $|\boldsymbol{v}_d|/v_{A,0} \approx 0.05$ --- nearly an order of magnitude smaller than in the $\xi = 10^{-2}$ run. This reflects the underlying scaling $\nu_{ni} = \gamma_d \rho_i$: with $\rho_i$ ten times larger than in the $\xi=10^{-2}$ case, the drag-mediated coupling between species is correspondingly tighter throughout the volume, and only modest velocity differences develop even at reconnection sites and around vigorously coalescing plasmoids. The combination of Figs.~\ref{fig:CS_xi} and \ref{fig:CS_drift} therefore confirms that the $\xi = 10^{-1}$ run operates predominantly in the coupling regime, with ion-neutral decoupling effects manifesting primarily through the modified inner-layer structure rather than through bulk dynamical decoupling.

\paragraph{Fully ionized baseline ($\xi = 1$):}
For reference, the fully ionized run reproduces the canonical single-fluid plasmoid-reconnection scenario \citep{Bhattacharjee2009,Huang_Bhattacharjee2010}. In the absence of a neutral fluid, the current sheet undergoes a prolonged quasi-Sweet--Parker thinning phase lasting until $t \approx 20\,t_{A,i}$, during which the global root mean square velocity $v_{\rm rms}$ remains below $10^{-3}\,v_{A,0}$ (Fig.~\ref{fig:energy}, top row). This onset delay is distinct from the exponential growth rate of an individual tearing mode. Plasmoid formation then sets in explosively at $t \approx 30\,t_{A,i}$, and by $t = 37\,t_{A,i}$ (Fig.~\ref{fig:CS_xi1}) the sheet has reorganized into a hierarchical chain dominated by a single macroscopic ``monster'' plasmoid near $x = L_x/2$, flanked by two intermediate secondary plasmoids. The monster plasmoid exhibits the characteristic wound-up, spiral substructure produced by successive coalescence events, with localized field amplification $|B|/B_0 \approx 1.5$--$2$ inside the trapped flux bundles and thin secondary current sheets bracketing each plasmoid.

\begin{figure*}
\centering
\includegraphics[width=0.99\linewidth]{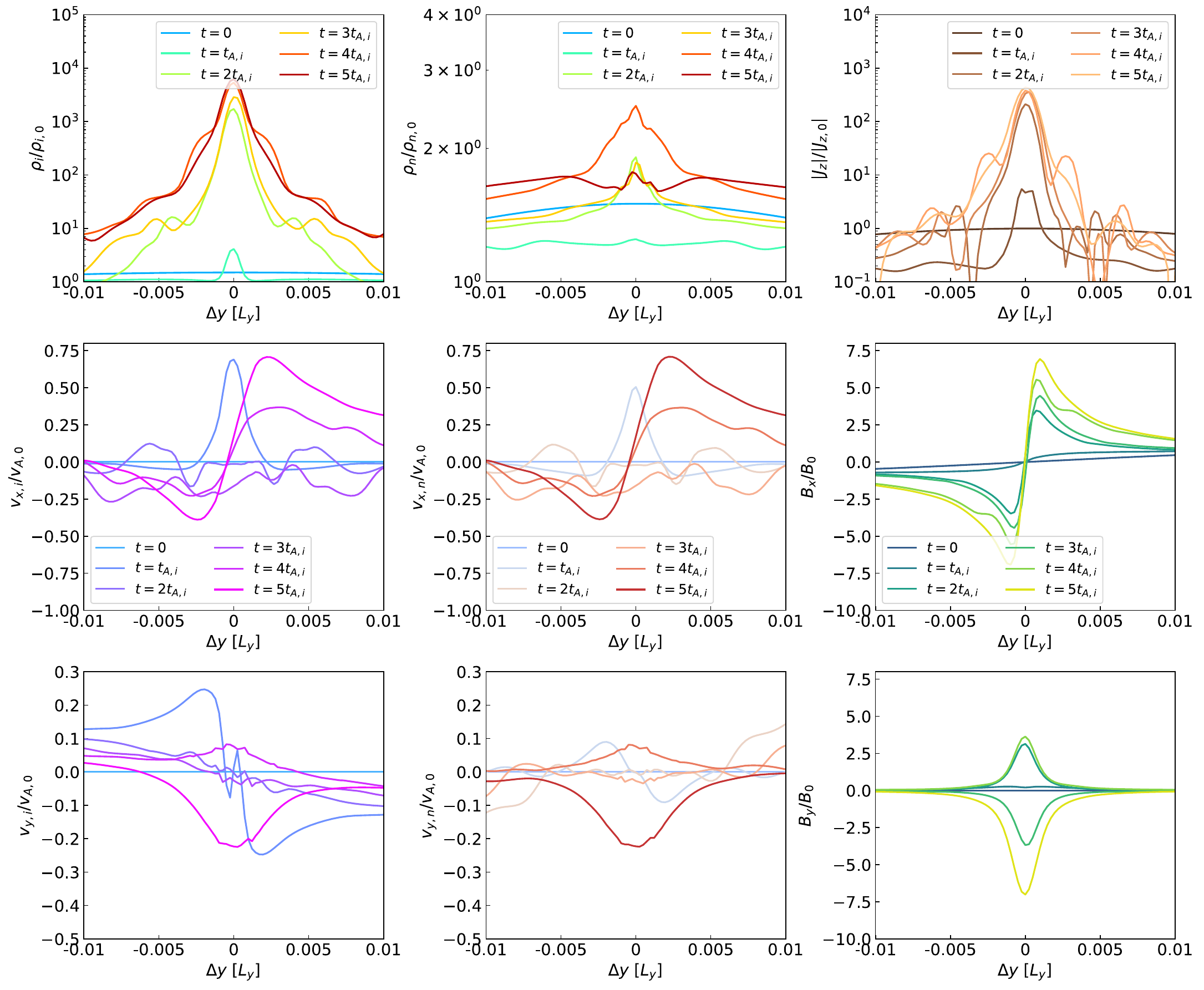}
        \caption{Temporal evolution of the one-dimensional transverse profiles across the primary central plasmoid for the run with initial ionization fractions $\xi = 10^{-2}$. The cross-sectional slices are extracted by continuously tracking the real-time position of the plasmoid centroid. The horizontal axis ($\Delta y$) represents the relative distance across the plasmoid centroid in the inflow direction. The $3 \times 3$ panels display the following physical quantities: normalized ion density ($\rho_i/\rho_{i,0}$, top left), neutral density ($\rho_n/\rho_{n,0}$, top middle), and the magnitude of the out-of-plane current density ($|J_z|/|J_{z,0}|$ on a logarithmic scale, top right); ion and neutral velocity components along the outflow direction ($v_{x,i}/v_{A,0}$ and $v_{x,n}/v_{A,0}$, middle left and center) alongside the reconnecting magnetic field ($B_x/B_0$, middle right); and the velocity components in the inflow direction ($v_{y,i}/v_{A,0}$ and $v_{y,n}/v_{A,0}$, bottom left and center) alongside the reconnected magnetic field component ($B_y/B_0$, bottom right). The different colored lines denote progressive snapshots from $t=0$ to $t=5 t_{A,i}$.}
    \label{fig:plasmoid_profile_xi001}
\end{figure*}

\begin{figure*}
\centering
\includegraphics[width=0.99\linewidth]{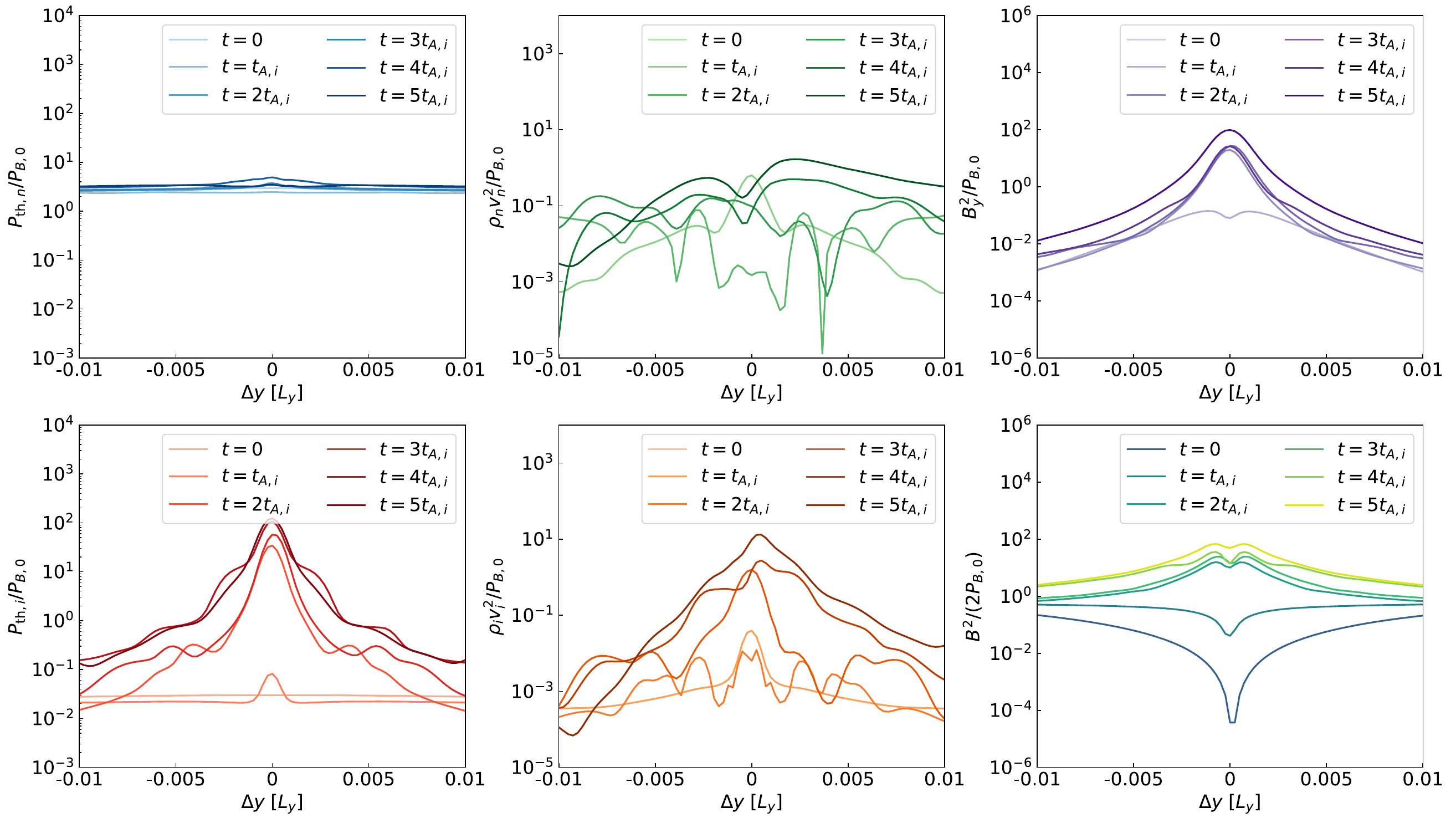}
        \caption{Temporal evolution of the one-dimensional transverse profiles across the primary central plasmoid the run with initial ionization fractions $\xi = 10^{-2}$. The cross-sectional slices are extracted by continuously tracking the plasmoid centroid's position in real time. The horizontal axis ($\Delta y$) represents the relative distance across the plasmoid centroid in the inflow direction. The $3 \times 3$ panels display the following physical quantities: neutral thermal pressure ($P_{{\rm th}, n}$, top left), neutral ram pressure ($\rho_nv_n^2$, top middle), and the squared reconnected field ($B_y^2$, top right); ion thermal pressure ($P_{{\rm th}, i}$, bottom left), ion ram pressure ($\rho_iv_i^2$, bottom middle), and the magnetic pressure ($B^2$, bottom right);  The different colored lines denote progressive snapshots from $t=0$ to $t=5 t_{A,i}$.}
    \label{fig:plasmoid_pressure_xi001}
\end{figure*}

\subsection{Internal Structure of the Primary Plasmoid: $\xi = 10^{-2}$}
\label{sec:plasmoid_xi001}

The internal structure of the primary plasmoid in the $\xi = 10^{-2}$ run (Fig.~\ref{fig:plasmoid_profile_xi001}) is qualitatively different from the transitional $\xi = 10^{-1}$ case. The plasmoid at $t = 5\,t_{A,i}$ is $\sim 16$ cells, smaller than the upstream decoupling length $\ell_{\rm dec} = 50$ cells and well below the $\xi = 10^{-1}$ plasmoid scale of $\sim\!160$ cells.

\subsubsection{Plasmoid density}
The density panels show the same two-stage evolution seen in the transitional $\xi = 10^{-1}$ case, but at much greater amplitude. By $t = t_{A,i}$, the current sheet has thinned substantially under the dominant ambipolar-driven flux compression. The ion fluid within the nascent sheet is sharply compressed into a narrow peak by the local magnetic-pressure gradient, while the neutrals --- feeling no direct magnetic force --- diffuse outward into a broad, low-amplitude profile. Beyond $t = t_{A,i}$, both densities continue to rise as reconnected flux accumulates in the plasmoid. The ion peak reaches $\rho_i/\rho_{i,0} \sim 3$--$5\times10^3$ by $t = 4$--$5\,t_{A,i}$, two to three orders of magnitude above the 
$\xi = 10^{-1}$ values. The neutral peak rises only to $\rho_n/\rho_{n,0} \approx 2$--$2.5$, over a much broader width, because the neutrals are dynamically coupled to the ions only at scales $\gtrsim \ell_{\rm dec}$ --- well above the plasmoid scale --- so the entire ion pile-up is filtered out of the neutral response. The resulting normalized ion-to-neutral density contrast within the plasmoid exceeds two orders of magnitude. This extreme pile-up is sustained by two effects: the ions are continuously delivered into the plasmoid by reconnection, while at sub-$\ell_{\rm dec}$ scales they cannot relax against the surrounding plasma through ion--neutral drag. Resistive reconnection across the surrounding separatrix is too slow during the early plasmoid evolution to open an efficient escape channel for the trapped ions.


\subsubsection{Current density and magnetic field strength}
The current density and field strength scale up correspondingly. Peak $|J_z|/|J_{z,0}|$ reaches a few $\times 10^2$, more than an order of magnitude above the $\xi = 10^{-1}$ values, and the reconnecting field is locally amplified to $|B_x|/B_0 \sim 5$--$7$ at the plasmoid boundaries (reaching $\sim 10$ in maximum). The reconnected component $B_y/B_0$ reaches local extrema in the range $\pm 3$ to $\pm 7$, again more than an order of magnitude larger than in the transitional case, reflecting the strong magnetic compression of the decoupled ion fluid against the inertial barrier of the quasi-stationary neutral background. As in the $\xi = 10^{-1}$ case, the sign of the $B_y$ peaks varies between snapshots because the centroid tracker re-identifies the deepest local minimum of $A_z$ at each output time: small-scale coalescence events --- which proceed continuously even though the global hierarchy is truncated --- shift the tracked structure to a physically distinct flux rope whose reconnected flux is wound in the opposite sense, producing an apparent sign change with no underlying topological reversal.

\subsubsection{Plasmoid kinetics}
The middle row of Fig.~\ref{fig:plasmoid_profile_xi001} shows that the outflow velocities $v_{x,i}$ and $v_{x,n}$ are similar, both exhibiting antisymmetric profiles peaking at $\sim 0.5$--$0.75\,v_{A,0}$ near the plasmoid edges, with $v_{x,n}$ only slightly lower in amplitude. This near-equality is not in conflict with the plasmoid scale lying below the upstream decoupling length $\ell_{\rm dec}$, because the relevant scale inside the plasmoid is the \emph{local} decoupling length, not the upstream value. The ion pile-up at the plasmoid core raises the local ionization fraction, contracting $\ell_{\rm dec} \propto \rho_i^{-1}$ to far below any resolvable scale. The two fluids are therefore tightly coupled \emph{inside the plasmoid} despite being decoupled in the dilute plasmoid peripheries.

The kinematic decoupling between the two species is visible instead in the inflow direction (bottom row), and it evolves with the maturity of the plasmoid. At early times ($t \lesssim 2\,t_{A,i}$), $v_{y,i}$ develops sharp central spikes localized within a few cells of the centroid that are absent from the much smoother $v_{y,n}$ profile. At this stage $\rho_i$ has not yet risen sufficiently to contract $\ell_{\rm dec}$ below the feature scale, and the sharp ion structures are filtered out of the neutral response by the same mechanism that produces the density contrast: at sub-$\ell_{\rm dec}$ scales the drag is too weak to entrain the neutrals into the localized ion motions. As the plasmoid grows and the central ion density approaches its peak value by $t \gtrsim 3\,t_{A,i}$, the local coupling progressively strengthens and $v_{y,n}$ begins to track $v_{y,i}$ more closely; by $t = 5\,t_{A,i}$ the two inflow profiles are visibly closer than at $t = t_{A,i}$.
\begin{figure*}
\centering
\includegraphics[width=0.99\linewidth]{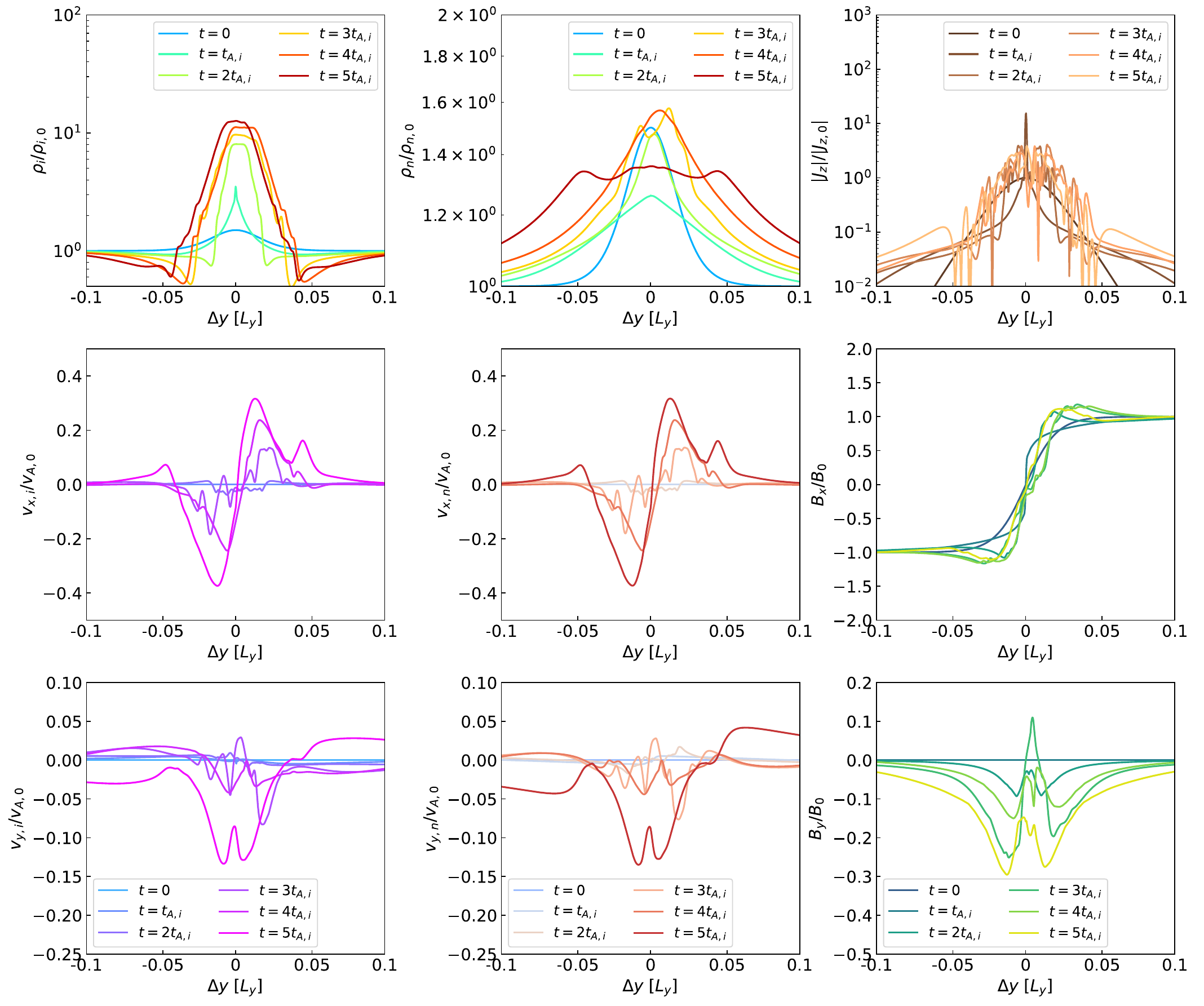}
        \caption{
        Same as Fig.~\ref{fig:plasmoid_profile_xi001}, but for initial ionization fractions of $\xi = 10^{-1}$.}
    \label{fig:plasmoid_profile_xi01}
\end{figure*}

\subsubsection{Pressure structure}
Fig.~\ref{fig:plasmoid_pressure_xi001} shows the transverse force budget of the $\xi=0.01$ run.  The
upstream decoupling length is large, $\ell_{\mathrm{dec}}\approx50$ cells ($\ell_{\mathrm{dec},i}\approx5$ cells), when the plasmoid core is still dilute, its internal structure lies below $\ell_{\mathrm{dec}}$ and the fluids are decoupled. As the core fills, however, the local ionization fraction rises and contracts the local $\ell_{\mathrm{dec}}\propto\rho_i^{-1}$ to below the scale of the internal features, so the two fluids re-couple within the dense core while remaining decoupled in the dilute plasmoid peripheries.
 
The early snapshots ($t\lesssim t_{A,i}$) again reproduce the Harris state, with flat ion and neutral thermal pressures ($P_{\mathrm{th}}/P_{B,0}\sim10^{-2}$ for the ions), negligible ram pressures, $B_y^2\approx0$, and a deep magnetic notch ($B^2/(2P_{B,0})\sim10^{-4.5}$). The plasmoid structure emerges between $t=2\,t_{A,i}$ and $t=4\,t_{A,i}$, during the early nonlinear evolution of this run, and the locally re-established coupling develops over the same interval.
 
The ion thermal pressure (bottom left) is flat through $t=t_{A,i}$, develops a thin central spike at $t=2\,t_{A,i}$ ($\sim\!10^{1}\,P_{B,0}$), and peaks at $t=3$--$4\,t_{A,i}$ at $\sim\!10^{2}\,P_{B,0}$ within a core a few cells across, following the $\rho_i/\rho_{i,0}\sim3$--$5\times10^{3}$ pile-up; by $t=5\,t_{A,i}$ the tip saturates and the shoulders broaden. The ion ram pressure (bottom middle) reaches a comparable peak, $\rho_i v_i^2/P_{B,0}\sim10^{1}$ at $t=4\,t_{A,i}$. Both profiles retain fine structure characteristic of an evolving driven layer. Continued ion accumulation accompanies limited mass removal through the surrounding current layers and weak neutral inertial feedback.

The magnetic terms confirm the non-static character of the balance. The reconnected $B_y^2$ (top right) builds a central peak rising monotonically to $\sim\!10^{2}\,P_{B,0}$ by $t=5\,t_{A,i}$. The initial central magnetic-pressure minimum (bottom right) progressively develops into a maximum of $\sim\!10^{1}$--$10^{2}\,P_{B,0}$ by $t=4$--$5\,t_{A,i}$, exceeding the upstream value of order unity. The decoupled ions drag the reconnecting field into the contracting core faster than it can slip outward, leaving the core as the site of the highest ion thermal pressure, ram pressure, and magnetic pressure simultaneously.
 
In the dilute upstream, where the local $\ell_{\mathrm{dec}}$ remains $\sim\!50$ cells, the neutral thermal pressure (top left) stays at its Harris value, and the neutral ram pressure (top middle) is suppressed by two to three orders of magnitude relative to the ions. Within the core, where the contracted local $\ell_{\mathrm{dec}}$ restores the drag coupling, the neutral ram pressure develops a central enhancement that grows with the plasmoid between $t=2$ and $5\,t_{A,i}$, tracking but never matching the ion terms. The $\xi=0.01$ plasmoid thus forms decoupled, because it is born smaller than the upstream $\ell_{\mathrm{dec}}$, and re-couples its
neutrals from the core outward as the pile-up contracts the local decoupling length, while remaining supported against the magnetic
stresses by a dynamic, reconnection-fed balance.
\begin{figure*}
\centering
\includegraphics[width=0.99\linewidth]{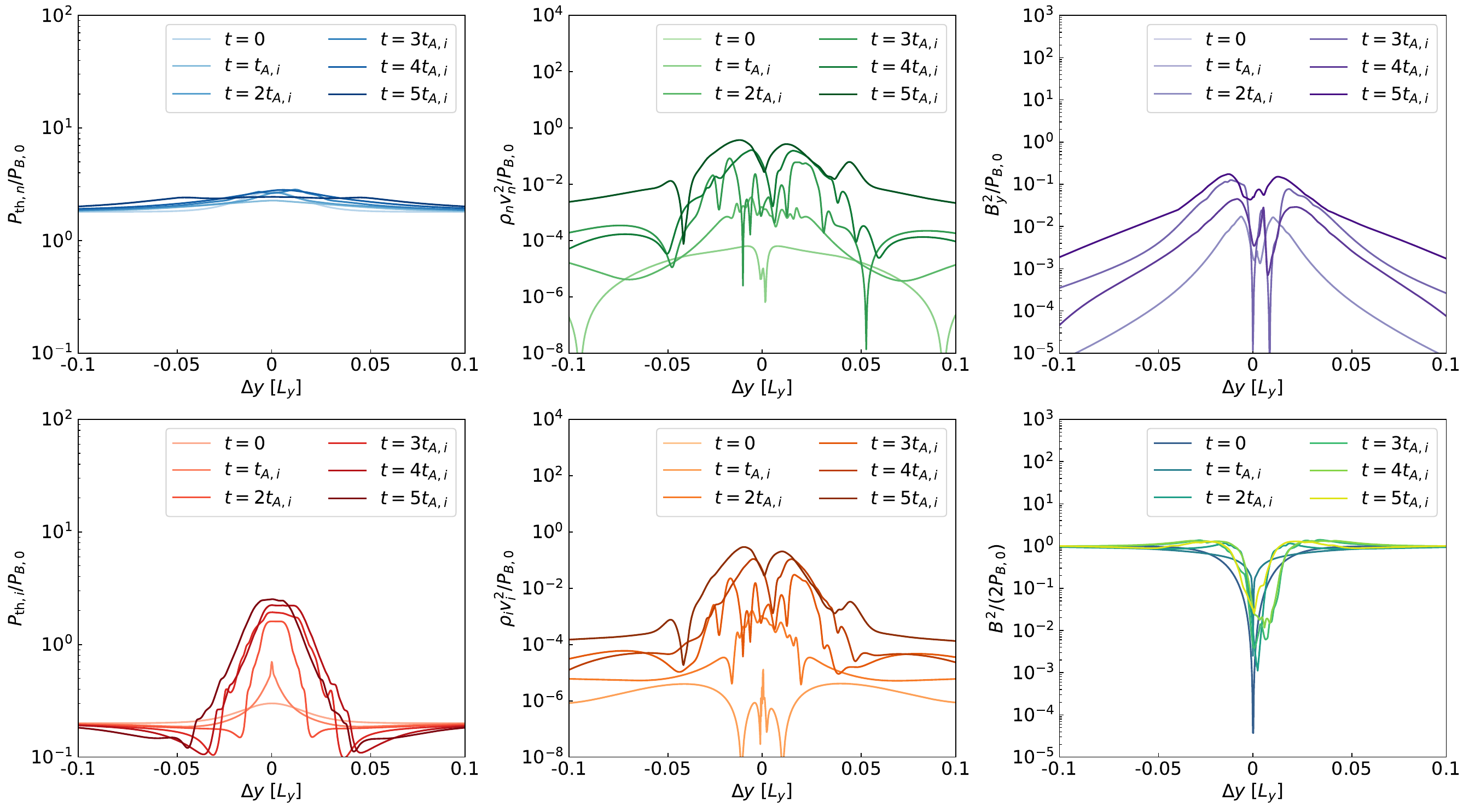}
        \caption{Same as Fig.~\ref{fig:plasmoid_pressure_xi001}, but for initial ionization fractions of $\xi = 10^{-1}$.}
    \label{fig:plasmoid_pressure_xi01}
\end{figure*}

\subsection{Internal Structure of the Primary Plasmoid: $\xi = 10^{-1}$}
\label{sec:plasmoid_xi01}
Figs.~\ref{fig:plasmoid_profile_xi01} and \ref{fig:plasmoid_profile_xi001} show the internal kinematic and magnetic profiles of the principal plasmoid, emerging at an early time but isolated, in each partially ionized run. The transverse cuts are extracted at the instantaneous centroid of the central plasmoid (identified as the deepest local minimum of $A_z$) and span a window $\Delta y \in [-0.1, 0.1]$ for $\xi = 10^{-1}$ and the much narrower window $\Delta y \in [-0.01, 0.01]$ for $\xi = 10^{-2}$.

\subsubsection{Plasmoid density}
For $\xi = 10^{-1}$ (Fig.~\ref{fig:plasmoid_profile_xi01}), the density panels reveal a two-stage evolution. At $t = t_{A,i}$, the current sheet has thinned under the combined action of Ohmic resistivity and ambipolar-driven flux compression, with the latter dominating the configuration of the inner layer ($\eta_\text{AD}/\eta \approx 68$). The resulting inner-layer width $\delta_{\rm AD} \approx 12$ cells is established more than an order of magnitude faster --- in $\sim t_{A,i}$ here, against $\sim 20t_{A,i}$ for purely Ohmic thinning in the fully ionized case. 

The density at the plasmoid centroid follows a non-monotonic trajectory in time. The initial Harris equilibrium, by construction, gives a density profile rising from $\rho/\rho_0 = 1$ at the upstream asymptote to $1.5$ at the sheet center (for $\beta= 2$); this profile is the baseline against which subsequent dynamical enhancement must be measured. During the linear thinning phase, $t \lesssim t_{A,i}$, no reconnection outflows have yet developed: the $v_{x,i}$ and $v_{x,n}$ profiles (Fig.~\ref{fig:plasmoid_profile_xi01}, middle row) remain essentially flat through $t = t_{A,i}$, and the centroid evolution in this interval is driven by sheet thinning and ambipolar drift rather than by advective transport. Even so, the two species already respond in opposite senses. The ion density at the tracked centroid rises above the Harris peak as the ion fluid narrows under magnetic-pressure compression accompanying sheet thinning. The neutrals, which cannot be compressed by the magnetic pressure on sub-$\ell_{\rm dec}$ scales, slip outward through the ions via ambipolar drift; the neutral profile broadens and the centroid value drops below the Harris peak of $1.5$. This early opposite-sign response is itself a signature of decoupling --- the ions follow the contracting magnetic structure on its own intrinsic scale, while the neutrals see only their own pressure gradient.

Beyond $t \approx t_{A,i}$, reconnected flux continues to accumulate and successive coalescence events feed the plasmoid, but with different amplitudes for the two species. The neutral peak reaches only $\rho_n/\rho_{n,0} \approx 1.6$, marginally above the initial Harris value of $1.5$ and comparable to the fully ionized case. The ions, by contrast, climb to $\rho_i/\rho_{i,0} \approx 12$, fully an order of magnitude beyond the Harris baseline. This $\sim 8\times$ ion-side dynamical excess, against the $\sim 1.07\times$ neutral-side excess in the same plasmoid, is the diagnostic signature of sub-$\ell_{\rm dec}$ pile-up: at scales below the decoupling length, the ions respond to the local magnetic-pressure of the plasmoid on its own, while the neutrals can be advected only on scales above $\ell_{\rm dec}$ and therefore remain locked to the upstream equilibrium response. The neutral density profile is correspondingly broad and weakly perturbed compared to the sharply peaked ion distribution.

\subsubsection{Current density and magnetic field strength}
The reconnecting field $B_x/B_0$ does not execute a simple monotonic reversal; instead, it exhibits a flat plateau outside the plasmoid ($|B_x|/B_0 \approx 1$ for $|\Delta y| > 0.04$), steep transitions across the wrapping current sheets at $\Delta y \approx 
\pm 0.04$, and a smoother variation through the plasmoid interior. This is the spatial signature of a magnetic plasmoid in 2D reconnection: the gradient of $B_x$ --- and hence the contribution to $J_z$ --- is sharply concentrated at the wrapping layers, where the field topology transitions from open upstream loops to closed plasmoid flux surfaces, while the interior carries a smoothly distributed current of much lower amplitude. Ohmic dissipation $\eta J_z^2$ is therefore localized at the boundary current sheets rather than in the plasmoid core. 

The reconnected field component $B_y/B_0$ displays the expected bipolar (``quadrupolar'') pattern, with peaks of magnitude $\approx 0.3$ on either side of the centroid and a near-zero crossing at $\Delta y = 0$ where the field lines close to form the magnetic plasmoid. The growth of $|B_y|$ from $t = t_{A,i}$ to $t = 5\,t_{A,i}$ traces the accumulation of reconnected flux within the plasmoid. The sign of the $B_y$ peaks can flip between snapshots because the centroid tracker re-identifies the deepest local minimum of $A_z$ at every output time: following a coalescence event, the new ``central'' plasmoid is in general a physically distinct flux rope whose reconnected flux is wound in the opposite sense, so the apparent sign change reflects a relabeling of the tracked structure rather than a reversal of the underlying flux topology.

\subsubsection{Plasmoid kinetics}
The middle row of Fig.~\ref{fig:plasmoid_profile_xi01} shows that the outflow components $v_{x,i}$ and $v_{x,n}$ are nearly identical antisymmetric profiles, peaking at $\pm 0.3\,v_{A,0}$ near the plasmoid edges. On the plasmoid scale, the two fluids are tightly coupled by collisional drag and advect essentially as a single fluid. The order-of-magnitude density contrast between them is, therefore, not a kinematic effect --- the bulk flows are coherent --- but reflects how the two species respond to the magnetic pressure on scales below $\ell_{\rm dec}$. The bottom row shows the corresponding inflow signature. The inflow velocities $v_{y,i}$ and $v_{y,n}$ track each other closely, with bulk values $\approx -0.1\,v_{A,0}$ representing the global downward motion of the centroid in the simulation frame, modulated by a sharp central dip at $\Delta y = 0$. 

\subsubsection{Pressure structure}
Fig.~\ref{fig:plasmoid_pressure_xi01} shows the transverse force budget of the $\xi=0.1$ plasmoid, decomposed into the thermal ($P_{\mathrm{th}}$), ram ($\rho v^2$), and magnetic ($B^2$, with the reconnected component $B_y^2$ shown separately) pressures of each fluid over the $t=0\to5\,t_{A,i}$ sequence.
 
At $t=0$ the panels reproduce the Harris equilibrium: the ion thermal pressure (bottom left) is flat at $P_{\mathrm{th},i}/P_{B,0}\approx0.15$, the ram pressures (middle column) are negligible, and the magnetic pressure (bottom right) carries the single reversal notch at $\Delta y=0$ with $B_y^2\approx0$ (top right). The reconnected field grows: from $t=t_{A,i}$ the $B_y^2$ profile develops the double-humped structure, its peaks rising from $\sim\!10^{-4}$ to $\sim\!10^{-1}\,P_{B,0}$ near $\Delta y\approx\pm0.01$ by $t=5\,t_{A,i}$, and the magnetic-pressure core deepens in step to $B^2/(2P_{B,0})\sim10^{-4}$ before refilling slightly as the plasmoid broadens.
 
The ion thermal pressure does not evolve toward a peaked profile that offsets this growing core deficit. A narrow central maximum forms early ($t\sim t_{A,i}$, $\sim\!2\,P_{B,0}$) and then broadens into a flat-topped plateau of $\sim\!2$--$3\,P_{B,0}$ across $|\Delta y|\lesssim0.03$ by $t=3$--$5\,t_{A,i}$. The ion ram pressure rises on the same timescale into a fragmented, multi-peaked profile, reaching $\rho_i v_i^2/P_{B,0}\sim10^{0}$ in the wings, comparable to the thermal term. Comparable thermal and ram pressures accompany the evolving magnetic-pressure structure. Reconnection supplies the ion overdensity ($\rho_i/\rho_{i,0}\approx12$) during plasmoid growth, consistent with a dynamically evolving core.
 
The neutral fluid is partially entrained into the evolving plasmoid. The neutral ram pressure (top middle) rises by three to four orders of magnitude, from $\sim\!10^{-5}$ to $\sim\!10^{-1}\,P_{B,0}$, as the neutrals are entrained into the developing outflows; it remains roughly an order of magnitude below the ion ram pressure, a difference set by the sub-$\ell_{\mathrm{dec}}$ slip that drag cannot suppress. The neutral thermal pressure (top left) stays near its upstream value $P_{\mathrm{th},n}/P_{B,0}=(1-\xi)\beta\approx1.8$ throughout, developing only a slight central enhancement, since the neutrals share the bulk motion through drag but feel no direct magnetic compression. The $\xi=0.1$ plasmoid is therefore a predominantly single-fluid plasmoid in a dynamic, reconnection-fed balance, with a modest sub-$\ell_{\mathrm{dec}}$ ion--neutral offset superimposed.
\begin{figure*}
\centering
\includegraphics[width=0.99\linewidth]{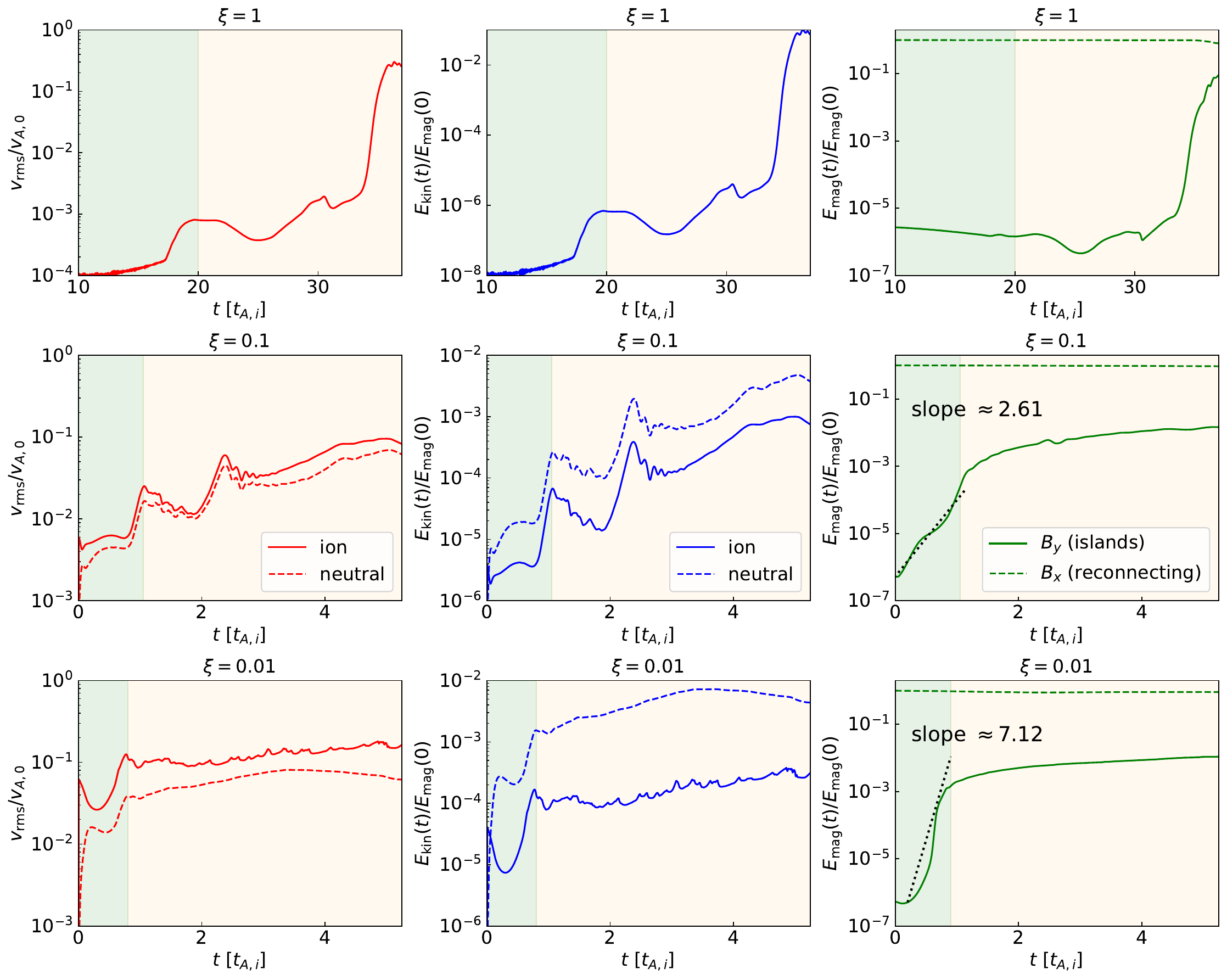}
        \caption{The three columns display, from left to right: the root-mean-square velocity ($v_{\rm rms}$) normalized by the upstream total Alfv\'en speed ($v_{A,0}$); the kinetic energy ($E_{\rm kin} = \frac{1}{2}\rho v^2$) normalized by the initial total magnetic energy $E_{\rm mag}(0)$; and the magnetic energy components ($E_{\rm mag} = B^2/8\pi$) normalized by $E_{\rm mag}(0)$. In the partially ionized cases ($\xi < 1$), solid and dashed lines in the left and middle columns denote the ion and neutral fluids, respectively. In the right column, the solid green lines represent the energy of the reconnected magnetic field component ($B_y$, corresponding to magnetic plasmoids), while the dashed green lines track the reconnecting field component ($B_x$). The background shading delineates the linear growth stage (light green) and the subsequent non-linear stage (pale yellow). The black dotted lines in the right panels indicate the exponential fits for the linear growth phase. Note the different temporal scales on the horizontal axes.}
    \label{fig:energy}
\end{figure*}

\subsection{Linear Growth and Two-Fluid Energy Partition}

Fig.~\ref{fig:energy} traces the global root-mean-square velocity $v_{\rm rms}/v_{A,0}$, the kinetic energy
$E_{\rm kin}(t)/E_{\rm mag}(0)$, and the magnetic-energy components $E_{\rm mag}(t)/E_{\rm mag}(0)$ for the three runs. In the right column, we decompose the magnetic energy into the reconnecting component $E_{B_x}$ (dashed) and the reconnected component $E_{B_y}$ (solid). Throughout the partially ionized runs, the ion (solid) and neutral (dashed) fluids are tracked separately in the left and middle columns. The dashed green curves in the right column show that the upstream reconnecting field $B_x$ remains essentially unchanged over the duration of all three simulations. This suggests that the domain is large enough to avoid global flux exhaustion and that all dynamical features reported here arise within a quasi-stationary upstream environment. A resolution-convergence study is presented in Appendix Fig.~\ref{fig:energy_res}.

As shown in Fig.~\ref{fig:energy}, the reconnected magnetic energy $E_{B_y}$ increases rapidly before the nonlinear plasmoid-dominated stage (light-green shading). However, $E_{B_y}$ is a global quantity that contains the imposed seed perturbation, multiple competing Fourier modes, and changes in the background current sheet. Its exponential amplification therefore should not be identified directly with the linear tearing growth rate.

\subsubsection{Tearing-mode growth}
We instead determine the tearing growth from the Fourier amplitude of the unseeded $kL=40\pi$ mode, as described in Appendix~\ref{sec:tearing_estimates}. The fitted amplitude growth rates are
\begin{equation}
\gamma_{40\pi}^{\rm fit}t_{A,0}
\simeq 0.39,\;18.4,\;102.9
\end{equation}
for $\xi=1$, $10^{-1}$, and $10^{-2}$, respectively. For the measured configured sheet widths, the corresponding theoretical fixed-mode tearing estimates are
$\gamma_{40\pi}^{\rm th}t_{A,0}\simeq2.0$, $37.5$, and $118.5$.

Both the measured and theoretical modal rates increase strongly as the configured sheet becomes thinner. In the coupled regime, the effective inertia approaches the total density and the standard single-fluid tearing scaling is recovered, while intermediate coupling can instead introduce frictional stabilization \citep{Zweibel1989,Pucci2020,2020JPlPh..86f5301P}. In our simulations, the important effect is that ambipolar transport produces substantially thinner current sheets before tearing develops, which subsequently support faster modal growth.

\begin{figure*}
\centering
\includegraphics[width=0.99\linewidth]{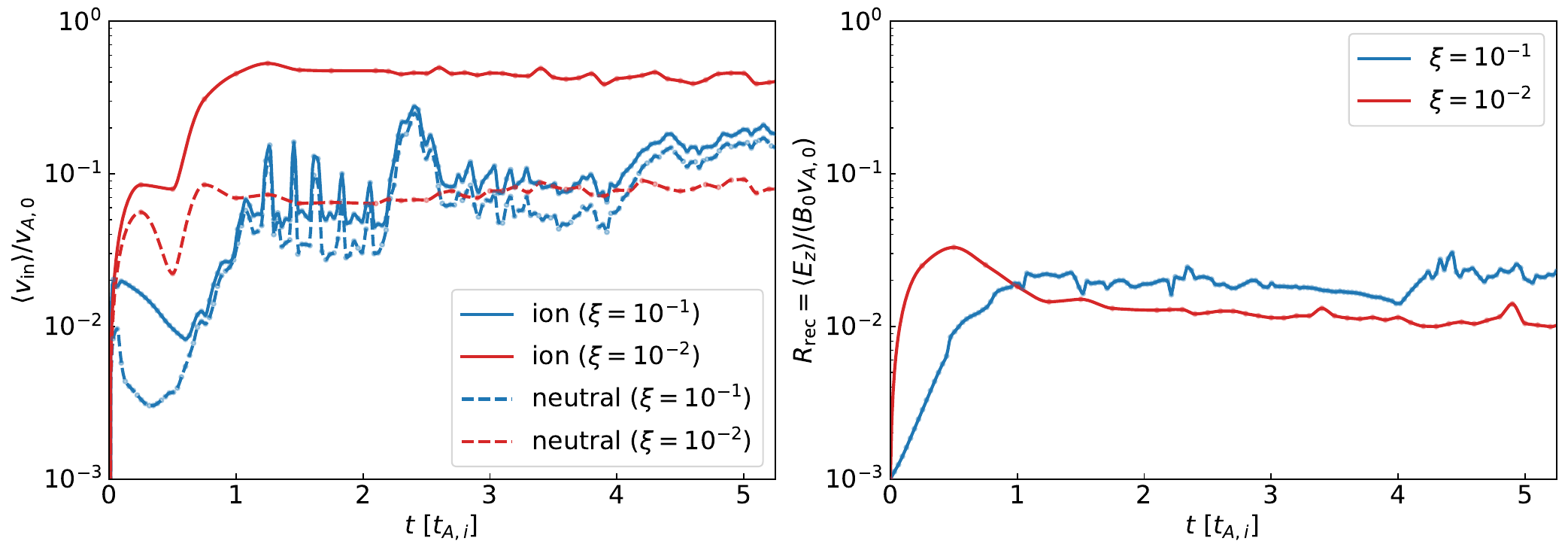}
        \caption{Time evolution of the inflow speed and the reconnection rate. \textit{Left:} the maximum inflow speed averaged over the active reconnection sites, normalized to the total Alfv\'en speed, shown separately for the ion (solid) and neutral (dashed) fluids. \textit{Right:} the reconnection rate $R_{\rm rec}=\langle E_z\rangle/(B_0 v_{A,0})$ defined from the out-of-plane electric field at the same sites.}
    \label{fig:rate}
\end{figure*}

\subsubsection{Ion--neutral decoupling}
The middle column reveals an asymmetric partition of kinetic energy between the two fluids in the nonlinear phase. For $\xi = 10^{-2}$, by contrast, $E_{\rm kin,n}/E_{\rm kin,i} \sim 10^2$: the neutrals overwhelmingly dominate the kinetic energy budget despite moving at lower velocities than the ions. The corresponding $v_{\rm rms}$ panel shows that the ion velocity saturates quickly at $v_{\rm rms,i}/v_{A,0} \approx 0.1$ from $t \gtrsim t_{A,i}$ onward, while the neutral $v_{\rm rms,n}/v_{A,0} \approx 0.03$ rises more gradually. The decoupled ion velocity is therefore $\sim\!3$ times larger than the neutral velocity in the most weakly ionized run.
For $\xi = 10^{-1}$, the neutral kinetic energy slightly exceeds the ion kinetic energy ($E_{\rm kin,n}/E_{\rm kin,i} \sim 2$) by $t = 5\,t_{A,i}$. 

\subsection{Reconnection Rate}

Quantifying the reconnection rate in our setup requires care. With periodic boundaries, the plasmoids cannot leave the domain, so the reconnected flux is not advected out but accumulates in the sheet and is removed only through Ohmic dissipation. We therefore measure the rate locally at the active reconnection sites, identified as the local minima of the vector potential $A_z$ along the sheet midplane (Fig.~\ref{fig:rate}). We calculate the maximum transverse inflow velocity averaged over these sites, $\langle v_{\rm in}\rangle/v_{A,0}$, separately for the ions and neutrals, and the reconnection rate defined from the peak out-of-plane electric field $E_z = -\partial_t A_z$,
\begin{equation}
R_{\rm rec}(t) = \frac{\langle E_z\rangle(t)}{B_0\,v_{A,0}},
\end{equation}
at the same reconnection sites. Since $E_z$ directly measures the rate at which flux is reconnected, it is insensitive to the non-reconnection motions---plasmoid advection and coalescence-driven flows---that contaminate the inflow-velocity estimate, and we adopt it as the reconnection rate throughout.

In the strongly decoupled $\xi = 10^{-2}$ run, the ion inflow velocity rises sharply to $\langle v_{\rm in}\rangle\sim0.5\,v_{A,0}$ within $t\lesssim t_{A,i}$ and remains high, far above the neutral inflow velocity ($\sim0.1\,v_{A,0}$, dashed red): the ambipolar drift draws the ions rapidly into the sheet while the neutrals lag behind. The reconnection electric field, by contrast, peaks at only $R_{\rm rec}\approx0.035$ near $t\approx0.5\,t_{A,i}$---during the initial linear tearing stage---and then relaxes to a quasi-steady $R_{\rm rec}\approx0.01$. The rate seen in the $\xi = 10^{-2}$ case is achieved without the formation of large-scale plasmoids; coalescence is slowed at sub-$\ell_{\rm dec}$ scales and the current sheet is instead populated by a dense, steady-state ensemble of small plasmoids (see Fig.~\ref{fig:CS_Jz}).
At the active reconnection sites, the electric field is dominated by the Ohmic term, $E_z\simeq\eta J_z$. The coexistence of rapid ion inflow, strong ion accumulation, and a modest reconnection electric field shows that ion transport into the layer and magnetic-flux reconnection proceed at different rates. The local relation $E_z\simeq\eta J_z$ identifies the topology-changing term in this model; it does not by itself establish which process controls the reconnection rate, since the current-sheet geometry and ion transport balance also influence $J_z$.
 
The $\xi = 10^{-1}$ run shows the same qualitative ordering but with a smaller ion--neutral inflow contrast, and its $E_z$-based rate rises monotonically to a quasi-steady $R_{\rm rec}\approx0.02$. The two runs thus remain in the nonlinear plasmoid regime, with rates of a few percent of $v_{A,0}$ \citep{Uzdensky2010, Loureiro2012, Comisso2016}, but follow distinct temporal paths: $R_{\rm rec}$ in the $\xi = 10^{-1}$ run climbs gradually to a higher steady value $\approx0.02$ at $t\approx1.5\,t_{A,i}$ and further increase to $\approx0.035$ at $t\approx4\,t_{A,i}$ when apparent coalescence occurs (see Fig.~\ref{fig:CS_Jz}).  Since $E_z\simeq\eta J_z$ at the reconnection sites and $\eta$, $B_0$, and $v_{A,0}$ are identical in the two runs, the slightly lower steady rate of the $\xi=10^{-2}$ run reflects a lower mean $J_z$ at its reconnection sites: the strongly fragmented chain distributes the reconnection over a larger number of weaker $X$-points, lowering the ensemble-averaged $E_z$ relative to the fewer, stronger sites of the $\xi=10^{-1}$ run. 
 
\section{Discussion}
\label{sec:discussion}

\subsection{Astrophysical Implications}
The weakly ionized regime is important for the ISM \citep{1992pavi.book.....S,Draine_McKee1993, Ferriere2001,Leake_Arber2006,2008ApJ...677.1151L,2010ApJ...720.1612M,2012ApJ...757..154L,2021ApJ...915...67H,2022MNRAS.510.4952L}, the solar chromosphere, and protoplanetary disks, where reconnecting layers can enter the ambipolar diffusion range, and ions and neutrals no longer respond identically to the magnetic field. We caution that our isothermal simulations are 2D and neglect ionization-recombination chemistry; the implications discussed below should therefore be further tested.

In the weakly ionized run, macroscopic ``monster'' plasmoids are suppressed, and the layer fragments into a dense chain of sub-scale plasmoids threaded by numerous thin current sheets (see Fig.~\ref{fig:plasmoid_stats}). If similar fragmentation occurs in chromospheric or prominence current sheets, magnetic energy release may be distributed over many short-lived, fine-scale events. Our results suggest that neutral decoupling shifts the nonlinear plasmoid evolution toward a more fragmented, multi-threaded layer, rather than simply rescaling the fully ionized plasmoid hierarchy.

The ion pile-up at sub-$\ell_{\mathrm{dec}}$ scales embedded in a smoother neutral envelope follows from the species-dependent response to magnetic stresses at a fixed scale, and we expect it to persist beyond the 3D plasmoid geometry studied here. The nonlinear geometry will likely be richer in 3D, where the plasmoids correspond to flux ropes that can kink, reconnect, fragment, or couple to ambient turbulence. Quantitative comparison with observations will require models that include ionization balance and guide-field geometry. Overall, our results suggest that weak ionization changes the pathway by which reconnection proceeds: ambipolar drift promotes early current-sheet thinning and fragmentation, while ion--neutral decoupling imprints different small-scale structures on the ionized and neutral components.

The reconnection rate exhibits a regime-dependent temporal behavior: the strongly decoupled $\xi=10^{-2}$ run overshoots and relaxes to a lower steady rate, while the transitional $\xi=10^{-1}$ run rises monotonically to a higher value. In the $\xi=10^{-2}$ cases, the rapid ion inflow is not matched by an equally rapid reconnection of flux. This behavior is consistent with an ion-density pile-up of up to four orders of magnitude above its initial value, because rapid ion inflow supplies ions faster than they can be processed and expelled by reconnection.

\subsection{Comparison with earlier work}

n these simulations, ambipolar drift produces thinner current sheets that support faster growth of the common $kL=40\pi$ mode (Appendix~\ref{sec:tearing_estimates}). At a given sheet configuration, the effect of ion--neutral coupling on tearing is regime-dependent, and intermediate coupling can introduce frictional stabilization \citep{Pucci2020}. Ambipolar transport reshapes the sheet without itself changing magnetic connectivity (\S~\ref{sec:AD}), which is enabled by Ohmic resistivity. The faster ion inflow in the $\xi=10^{-2}$ run coexists with a lower late-time $E_z$-based reconnection  rate than in the $\xi=10^{-1}$ run (Fig.~\ref{fig:rate}).

A related sequence was found by \citet{2024ApJ...967..136T} in a self-forming, weakly ionized sheet with ionization--recombination: ambipolar thinning initially outpaces tearing, then stalls as ions decouple near the null and reconnection begins. Our simulations show a similar progression from ambipolar thinning to tearing in the decoupled regime. However, we initialize a Harris equilibrium and omit ionization--recombination, so their onset time is not directly applicable; the ambipolar singularity near the null \citep{Brandenburg_Zweibel1994} is regularized here solely by Ohmic resistivity.

Our model also differs from the three-fluid, five-moment study of \citet{2026arXiv260223683W}, which includes electron dynamics and finds thinning to the ion inertial scale $d_i$, followed by Hall-mediated fast reconnection. Our two-fluid MHD model contains neither Hall physics nor a kinetic inertial scale; its sheet thickness is instead set by the ambipolar configuration and Ohmic resistivity. The two studies therefore probe distinct MHD and kinetic regimes, whose connection remains an important topic for future work.

\subsection{Limitations}
\subsubsection{Neglect of ionization and recombination}


The ion accumulation and enhanced core ionization reported here characterize a model without explicit chemical source terms. In \citet{Leake2012}, excess ions supplied by decoupled inflow are removed through comparable recombination and outflow losses. Their weak Lundquist-number dependence is obtained despite the reconnection-site electric field remaining Ohmic, illustrating that the global reconnection rate can be controlled by the coupled transport and chemical balance rather than by the local non-ideal term alone \citep{VL1999}.

For illustrative reaction terms,
\begin{equation}
S_i=\zeta\rho_n-\alpha\rho_i^2,
\end{equation}
where $\zeta$ is the ionization rate and $\alpha$ is a mass-density-based recombination coefficient, the local ion-loss time is
$\tau_{\rm rec}=(\alpha\rho_i)^{-1}$.
Near chemical equilibrium and at fixed total density, small perturbations relax on a timescale
$\tau_{\rm chem}=(\zeta+2\alpha\rho \xi_{i,\rm eq})^{-1}$,
where $\xi_{i,\rm eq}$ is the equilibrium ionization fraction. This timescale describes chemistry-only relaxation about equilibrium and does not include dynamical transport.

The relevant dynamical timescales include the sheet-thinning time,
$|d\ln a/dt|^{-1}$, during current-sheet formation; the tearing time,
$\gamma^{-1}$, during linear growth; and the ion-supply, escape, compression, and residence times during the nonlinear evolution. The replacement of ions within the reconnection core depends on both the local mass inventory and transport across its boundary. For a moving control volume,
\begin{equation}
\frac{dM_i}{dt}
=
\dot M_{i,\rm in}
-
\dot M_{i,\rm out}
+
\int_V
\left(\zeta\rho_n-\alpha\rho_i^2\right)dV,
\end{equation}
where the mass fluxes are evaluated relative to the velocity of the control-volume boundary. Recombination could therefore reduce ion pile-up, replenish the neutral component, and modify the local ion--neutral coupling and pressure balance. Its importance, however, is determined by the competition between the chemical and dynamical timescales. The extreme ion overdensities obtained here should therefore be regarded as upper limits appropriate to the regime in which recombination is slow compared with the relevant local dynamical evolution.

\subsubsection{Periodic outflow boundaries}

We adopt periodic boundaries along the outflow direction, which could in principle allow reconnected flux to recirculate and influence the layer. This does not occur within the analyzed interval. The ion and neutral outflow speeds are comparable, $v_{x,i}\approx v_{x,n}\sim0.3$--$0.75\,v_{A,0}$ (Figs.~\ref{fig:plasmoid_profile_xi01} and \ref{fig:plasmoid_profile_xi001}), corresponding to a minimum recirculation time of $\gtrsim27\,t_{A,i}$. This greatly exceeds the analyzed duration of $5\,t_{A,i}$, so reconnected flux cannot traverse the domain and return to the reconnection region during the reported evolution.

Finally, we consider only three ionization fractions at fixed $\beta$, $\eta$, and $\gamma_{\rm d}$. A broader exploration of $(\xi,\beta,\eta,\gamma_{\rm d})$, together with three-dimensional simulations that permit additional instabilities and turbulence, is left to future work.

\section{Conclusion}
\label{sec:conclusion}

We have carried out 2D high-resolution ($16384\times4096$ cells) two-fluid (ion $+$ neutral) simulations of Harris-sheet reconnection spanning the fully ionized ($\xi=1$), transitional ($\xi=10^{-1}$), and ambipolar-diffusion (AD) dominated ($\xi=10^{-2}$) regimes at fixed Lundquist number $S=10^5$ and plasma beta $\beta=2$. The simulations follow sheet evolution and plasmoid formation into the nonlinear stage without ionization or recombination. Our major findings are as follows.
\begin{enumerate}
\item \textit{Plasmoid hierarchy varies across the sampled intervals.} In the fully ionized run, successive mergers produce a hierarchy dominated by a macroscopic ``monster'' plasmoid. In the $\xi=10^{-2}$ run, the reconnected flux remains distributed among many small plasmoids, and no comparably dominant structure appears during the analyzed evolution. Possible reasons include ion–neutral friction damping plasmoid motion and slowing the nonlinear dynamics of plasmoid merging. 

\item \textit{Ambipolar transport contributes to the early evolution of the current sheet.} The current sheet first thins rapidly under AD-mediated flux compression, more than an order of magnitude faster than the purely Ohmic Sweet--Parker thinning of the fully ionized case; sub-$\ell_{\mathrm{dec}}$ species differentiation then sets in. The fitted growth rate of the common $kL=40\pi$ mode increases as the measured configured sheet becomes thinner in these runs (Appendix~\ref{sec:tearing_estimates}). For $\xi=10^{-2}$, the reference scales satisfy $\delta_{{\rm SP},i}<\delta_{\rm AD}<\delta_{{\rm SP},0}$, with $\delta_{\rm AD}$ representing an ambipolar transport estimate rather than the measured sheet half-width.

\item \textit{Ions and neutrals re-couple inside plasmoids.} Below $\ell_{\mathrm{dec}}$ the ions are concentrated into dense plasmoids, reaching peak overdensities $\rho_i/\rho_{i,0}\approx10$ ($\xi=10^{-1}$) and $\sim3-5\times10^{3}$ ($\xi=10^{-2}$), while the neutral density remains comparatively smooth $\rho_n/\rho_{n,0} \approx 1.6$ ($\xi = 10^{-1}$) and $\sim 2-2.5$ ($\xi = 10^{-2}$). The pile-up raises the local ionization fraction by up to two orders of magnitude, driving plasmoid cores toward near-full ionization and contracting the local $\ell_{\mathrm{dec}}\propto\rho_i^{-1}$ below the plasmoid scale, consistent with stronger coupling in the cores than in the dilute peripheries. The peak values and inferred recoupling characterize the model with chemistry omitted.

\item \textit{Ion inflow and magnetic-flux reconnection exhibit distinct temporal evolution.} In the $\xi=10^{-2}$ run, the ion inflow reaches $\sim0.5v_{A,0}$, substantially exceeding the neutral inflow of $\sim0.1v_{A,0}$, while the reconnection rate decreases from $R_{\rm rec}\approx0.035$ to $\approx0.01$. In the $\xi=10^{-1}$ run, the rate increases to $\approx0.02$ and later reaches $\approx0.035$ during an apparent plasmoid coalescence event. The local non-ideal electric field is dominated by Ohmic dissipation, with $E_z\simeq\eta J_z$.
\end{enumerate}

\begin{acknowledgments}
  Y.H. thanks Elizabeth Tolman for helpful discussions. Y.H. is supported by the Sherman Fairchild Postdoctoral Fellowship at the California Institute of Technology. Y.H. acknowledges the support for this work provided by NASA through the NASA Hubble Fellowship grant No. HST-HF2-51557.001 awarded by the Space Telescope Science Institute, which is operated by the Association of Universities for Research in Astronomy, Incorporated, under NASA contract NAS5-26555. This work used SDSC Expanse CPU and NCSA Delta CPU through allocations PHY230032, PHY230033, PHY230091, PHY230105,  PHY230178, and PHY240183, from the Advanced Cyberinfrastructure Coordination Ecosystem: Services \& Support (ACCESS) program, which is supported by National Science Foundation grants \#2138259, \#2138286, \#2138307, \#2137603, and \#2138296. A.L. acknowledges the support of NSF grants AST 2307840 and NASA SEP864 80NSSC25K0067. S.X. acknowledges the support from the NASA ATP award 80NSSC24K0896. 
\end{acknowledgments}

%

\vspace{5mm}

\software{\texttt{AthenaK} \citep{Stone2024_AthenaK}
          }



\appendix
\section{Fully ionized case}
 \begin{figure*}[b]
\centering
\includegraphics[width=0.99\linewidth]{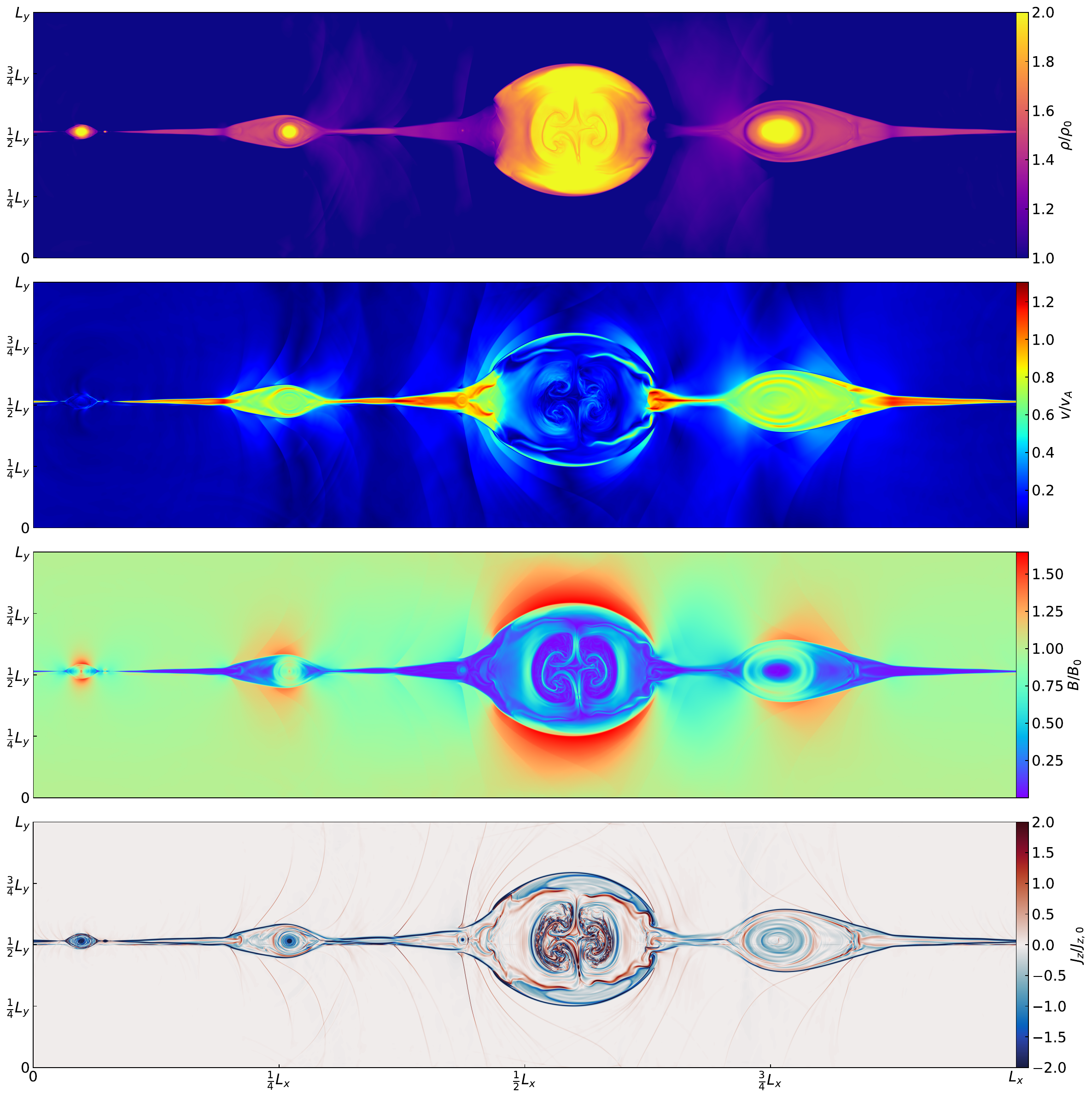}
        \caption{2D spatial distributions for the fully ionized ($\xi = 1$), single-fluid MHD limit in the fully developed non-linear regime ($t=37t_{A,i}$). From top to bottom, the panels display the normalized plasma density ($\rho/\rho_0$), velocity magnitude ($v/v_{A,0}$), magnetic field strength ($B/B_0$), and out-of-plane current density ($J_z/J_{z,0}$) across the entire computational domain. }
    \label{fig:CS_xi1}
\end{figure*}

Fig.~\ref{fig:CS_xi1} shows the 2D spatial distributions for the normalized plasma density, velocity magnitude, magnetic field strength, and out-of-plane current density in the fully ionized ($\xi = 1$), single-fluid MHD limit. In stark contrast to the highly fragmented, thinner current sheets observed in the partially ionized cases, this fully ionized scenario is dominated by the formation of a macroscopic "monster" plasmoid at the center, resulting from the successive coalescence of smaller magnetic plasmoids. The velocity map (second panel) clearly highlights the high-speed reconnection outflow jets feeding into these plasmoids. Furthermore, the high-resolution maps reveal highly intricate, wrapped internal substructures and nested cores within the primary central plasmoid across all physical quantities, reflecting the complex topological evolution and the dynamical "memory" of past merging events. 

\section{Plasmoid Statistics}
\label{sec:plasmoids}
We identify plasmoids using the in-plane flux function $A_z$, defined by $\mathbf{B}_{\perp}=\hat{\mathbf{z}}\times\nabla A_z$, so that $B_y=\partial_x A_z$, $B_x=-\partial_y A_z,$ and $A_z$ is constant along in-plane magnetic-field lines. Here, $x$ is the direction along the current sheet and $y$ is the direction across it, so that $B_x$ is the reconnecting field.

Because the plasmoids intersect the current-sheet midplane, we reconstruct the flux function along $y=0$ from
\begin{equation}
A_z(x,0)=
\int_{x_{\rm min}}^x
\left[B_y(x',0)-\langle B_y\rangle_x\right]
\,\mathrm{d}x',
\label{eq:fluxfn}
\end{equation}
where the numerically negligible residual mean of $B_y$ is removed to enforce periodicity and avoid an artificial linear drift in $A_z$.

For the magnetic polarity adopted here, O-points are local maxima of $A_z(x,0)$ and X-points are local minima. For an plasmoid centered at $\mathrm{O}_k$ and bounded by adjacent X-points $X_k^{-}$ and $X_k^{+}$, we define its reconnected flux as
\begin{equation}
\psi_k=
A_z(\mathrm{O}_k)
-\max\!\left[
A_z(X_k^{-}),A_z(X_k^{+})
\right],
\label{eq:plasmoid_flux}
\end{equation}
where the larger X-point value determines the outermost closed contour. Its along-sheet extent is
\begin{equation}
w_k=
\left|x(X_k^{+})-x(X_k^{-})\right|,
\end{equation}
with periodic wrapping included when necessary. Candidate extrema are identified using a prominence threshold proportional to the instantaneous flux range,
\begin{equation}
\Delta A_z=
\max_x A_z(x,0)-\min_x A_z(x,0).
\end{equation}
We count a plasmoid as a plasmoid when
\begin{equation}
\psi_k\geq f\Delta A_z,
\end{equation}
with $f=0.05$. Because the plasmoid hierarchy extends toward the resistive scale \citep{Uzdensky2010}, the absolute count depends on the threshold. We therefore verify that the qualitative trends are unchanged over the plateau of the count--threshold relation and cross-check the extrema using zero crossings of $B_y(x,0)=\partial_xA_z$.

 \begin{figure*}
\centering
\includegraphics[width=0.99\linewidth]{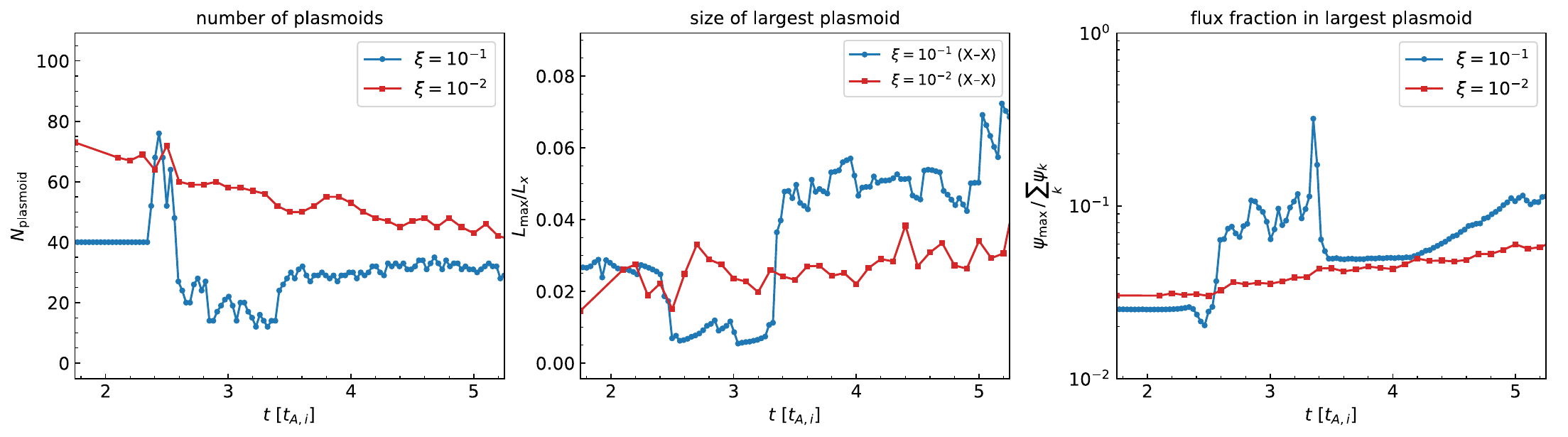}
\caption{Plasmoid statistics versus ion Alfv\'en times $t_{A,i}$ for $\xi=10^{-1}$ (blue) and $\xi=10^{-2}$ (red). \emph{Left:} number of plasmoids $N_{\rm plasmoid}$. \emph{Middle:} size of the largest plasmoid normalized by the sheet length, $L_{\rm max}/L_z$. \emph{Right:} fraction of the reconnected flux in the largest plasmoid, $\psi_{\rm max}/\sum_k\psi_k$. Plasmoids are identified from the midplane flux function $A_z(0,y)$ with flux content $\psi_k\ge0.05\,\Delta A_z$.}
\label{fig:plasmoid_stats}
\end{figure*}

Fig.~\ref{fig:plasmoid_stats} shows the plasmoid number, the along-sheet extent of the largest-flux plasmoid, and the fraction of identified plasmoid flux contained in that plasmoid. In the $\xi=10^{-1}$ case, the sheet initially contains about 40 plasmoids. The count briefly rises to $\sim70$ before dropping to $\sim15$--$25$, indicating rapid loss or coarsening of small plasmoids. Near $t\simeq3.3\,t_{A,i}$, the largest-plasmoid flux fraction transiently reaches $\sim0.3$, together with an increase in its extent, consistent with a major coalescence event and the temporary formation of a dominant plasmoid. Secondary tearing later regenerates smaller plasmoids, and the largest-plasmoid flux fraction decreases again. The evolution is therefore intermittent and strongly affected by plasmoid merging.

In contrast, the $\xi=10^{-2}$ case shows a smoother decline in plasmoid number without a comparable collapse. The largest plasmoid remains relatively small, with $w_{\max}/L_x\lesssim0.03$, and its flux fraction increases only gradually from $\sim0.02$ to $\sim0.06$. Thus, the reconnected flux remains distributed among many plasmoids, and the declining count is more consistent with gradual coarsening than with the formation of a single dominant plasmoid.

\section{Neutral density jump in the weakly ionized $\xi=10^{-2}$ run}

\begin{figure*}
\centering
\includegraphics[width=0.99\linewidth]{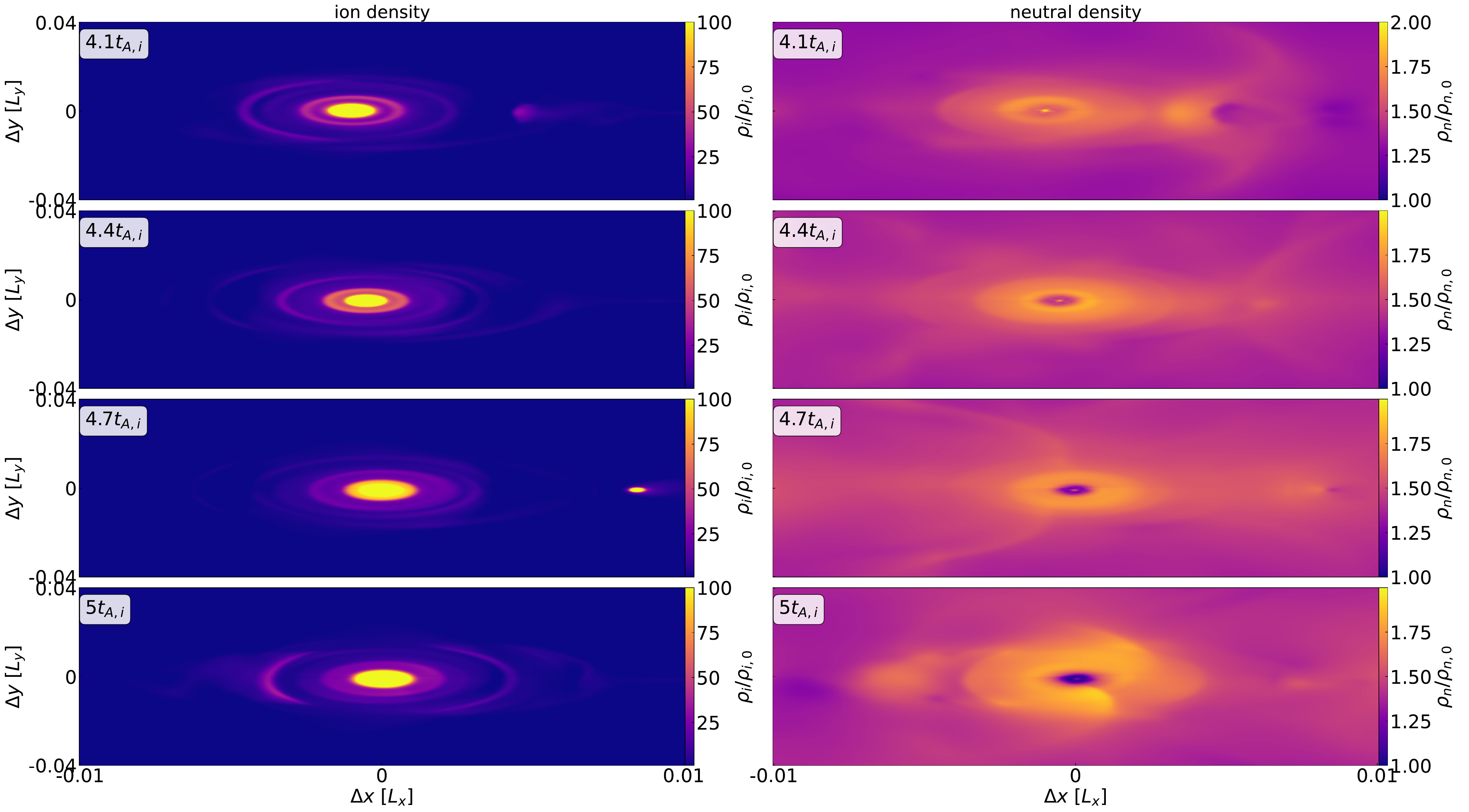}[hbtp!]
\caption{Ion (left) and neutral (right) density maps in the $\xi=10^{-2}$ run, shown at four times during the nonlinear phase ($t=4.1$, $4.4$, $4.7$, and $5\,t_{A,i}$, top to bottom). Each panel is a zoom on the tracked plasmoid, spanning $\Delta x\in[-0.01,0.01]\,L_x$ and $\Delta y\in[-0.04,0.04]\,L_y$.}
    \label{fig:CS_density_jump}
\end{figure*}

\begin{figure*}
\centering
\includegraphics[width=0.99\linewidth]{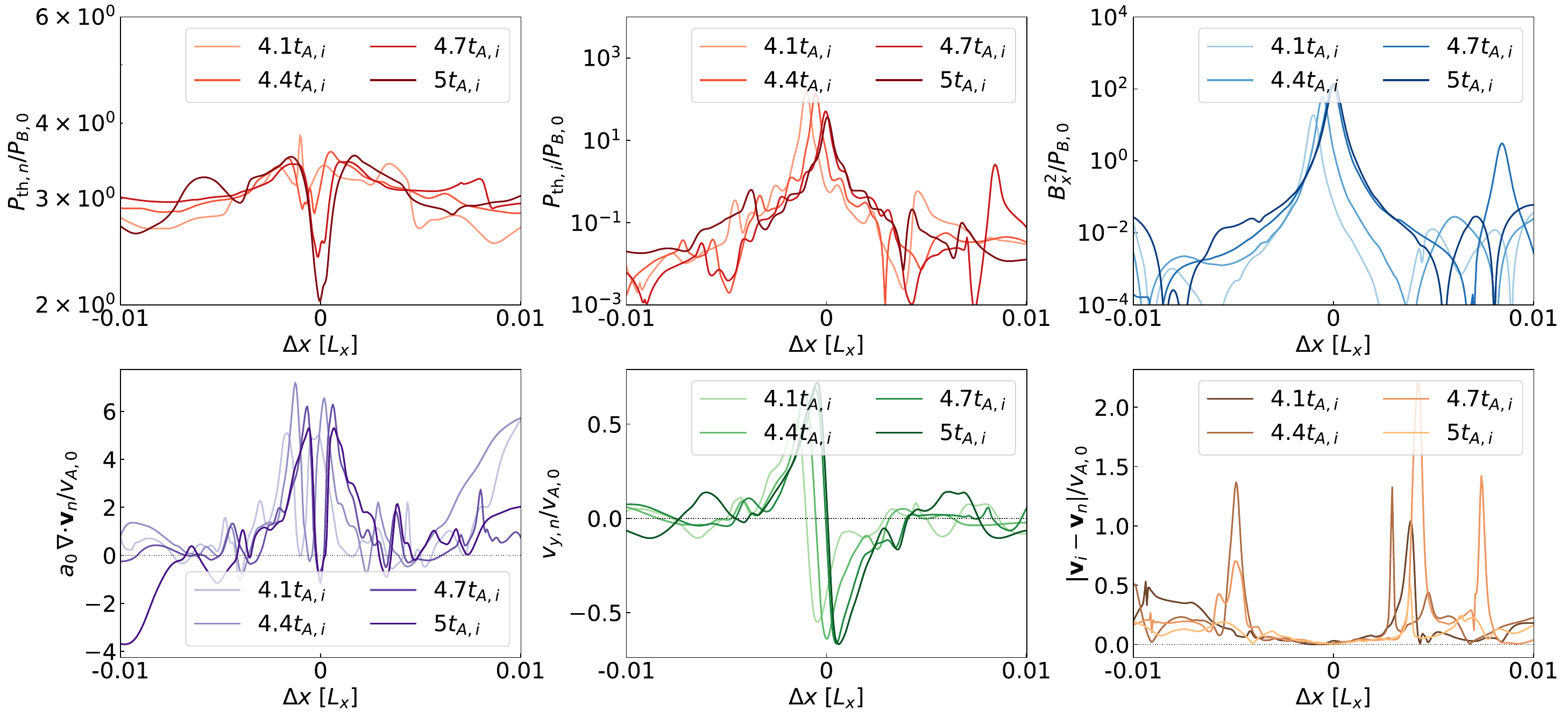}
        \caption{1D transverse profiles through the plasmoid, cut along the outflow direction ($\Delta x$) at the four times of Fig.~\ref{fig:CS_density_jump} ($t=4.1$, $4.4$, $4.7$, and $5\,t_{A,i}$; lighter to darker shades). \textit{Top row:} neutral thermal pressure $P_{\mathrm{th},n}/P_{B,0}$ (left), ion thermal pressure $P_{\mathrm{th},i}/P_{B,0}$ (middle), and the reconnecting- field $B_x^2/P_{B,0}$ (right). \textit{Bottom row:} the scaled neutral velocity divergence $a_0\,\nabla\!\cdot\!\boldsymbol{v}_n/v_{A,0}$ (left), the ion outflow velocity $v_{y,i}/v_{A,0}$ (middle), and the ion--neutral drift speed $|\boldsymbol{v}_i-\boldsymbol{v}_n|/v_{A,0}$ (right). }
    \label{fig:plot_pressure}
\end{figure*}

\begin{figure*}
\centering
\includegraphics[width=0.9\linewidth]{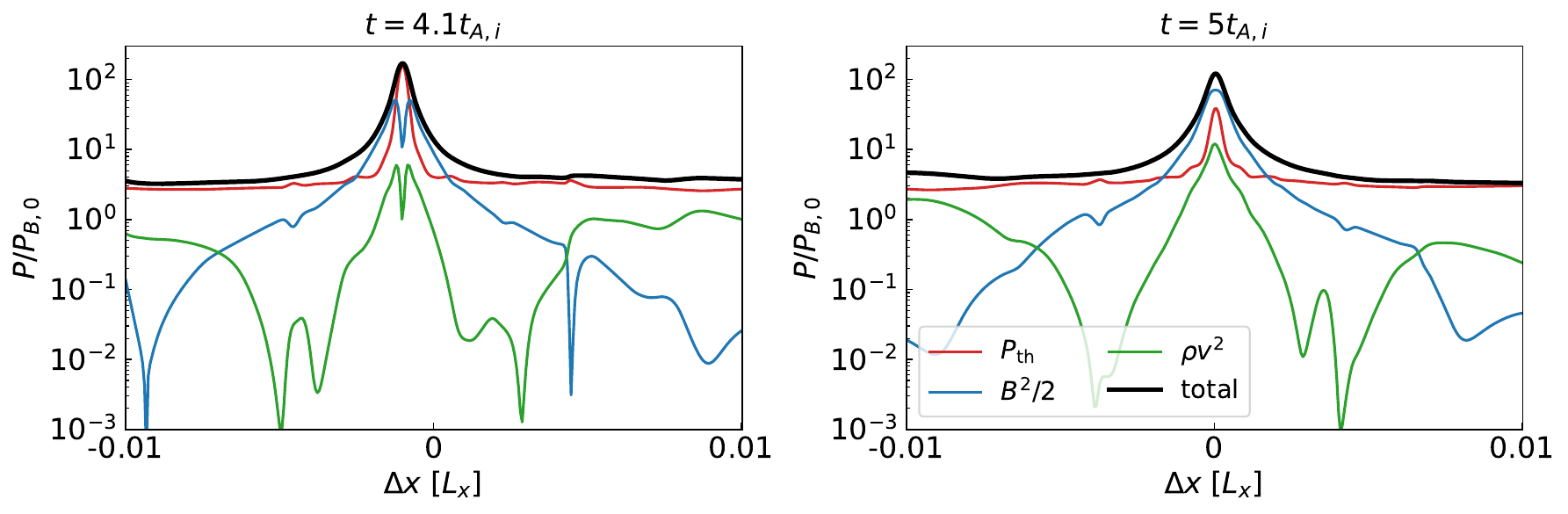}
\caption{Total pressure budget through the plasmoid presented in Fig.~\ref{fig:CS_density_jump}, cut along the outflow direction ($\Delta x$) at $t=4.1\,t_{A,i}$ (left) and $t=5\,t_{A,i}$ (right). Each curve is the sum of the ion and neutral contributions: thermal pressure $P_{\mathrm{th}}$ (red), magnetic pressure $B^2/2$ (blue), ram pressure $\rho v^2$ (green), and their total (black), all normalized to the upstream magnetic pressure $P_{B,0}$. }
    \label{fig:plot_pressure_total}
\end{figure*}

Fig.~\ref{fig:CS_density_jump} shows ion and neutral density maps of the tracked plasmoid in the $\xi=10^{-2}$ run at $t=4.1$, $4.4$, $4.7$, and $5t_{A,i}$. The ions form a sharply compressed, wound core with $\rho_i/\rho_{i,0}>100$, whereas the neutrals show only a smooth enhancement of about a factor of two. During relaxation, the ion core broadens and elongates but remains dense. By $t=5t_{A,i}$, the neutral density instead develops a central depletion, reaching $\rho_n/\rho_{n,0}\sim0.6$ at the centroid.

The outflow profiles in Fig.~\ref{fig:plot_pressure} show the same evolution. At $t=4.1t_{A,i}$, the central ion thermal pressure and normalized squared reconnecting field, $B_x^2/P_{B,0}$, peak at $\sim30$ and $\sim10^2$, respectively, and then decline as the compressed core and trapped magnetic flux relax. The neutral pressure remains much smaller, $\sim2$--$3.5P_{B,0}$, and develops a shallow central minimum. Antisymmetric neutral outflows reach $|v_{y,n}|\sim0.5$--$0.7v_{A,0}$, while $a_0\nabla\cdot\boldsymbol{v}_n/v_{A,0}\sim4$--$6$ indicates expansion. The ion--neutral drift is small at the centroid but reaches $\sim1$--$1.5v_{A,0}$ near the surrounding current sheets and outflow fronts, allowing the two fluids to relax at different rates.

Fig.~\ref{fig:plot_pressure_total} indicates that this expansion is pressure-driven. The wings remain near the upstream thermal pressure, $\sim3$--$4P_{B,0}$, while the core retains a strong total-pressure excess. Near maximum compression at $t=4.1t_{A,i}$, the peak is thermally dominated, with $P_{\mathrm{th}}\sim2\times10^2P_{B,0}$ and $B^2/2\sim1.5\times10^2P_{B,0}$. By $t=5t_{A,i}$, magnetic pressure dominates, with $B^2/2\sim10^2P_{B,0}$ and $P_{\mathrm{th}}\sim30$--$40P_{B,0}$. Meanwhile, the ram pressure in the wings grows and becomes smoother as the outflow develops.

This thermal-to-magnetic transition marks the shift from peak compression to relaxation. As coalescence-driven inflow weakens, expansion rapidly lowers the thermal pressure, whereas the wound magnetic flux decays more slowly. The ions therefore broaden while remaining magnetically confined and overdense. Neutrals, coupled only through drag, expand more freely on sub-$\ell_{\mathrm{dec}}$ scales, producing positive divergence and a central density depletion. The core retains evolving thermal, magnetic, and ram-pressure contributions throughout its relaxation.

\section{Resolution study}

\begin{figure*}[htbp!]
\centering
\includegraphics[width=0.99\linewidth]{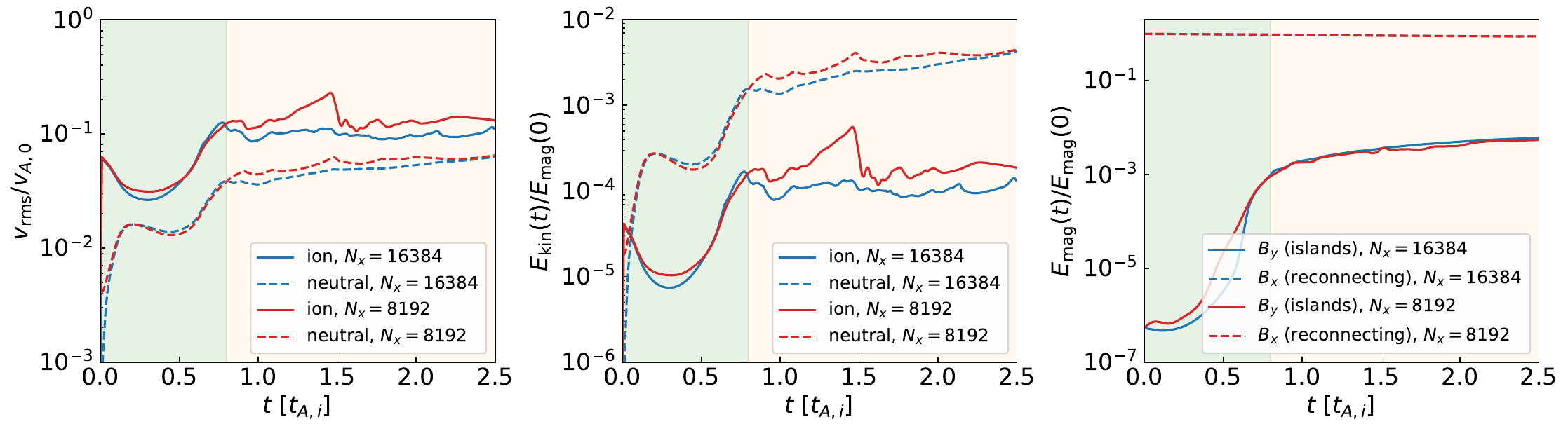}
        \caption{The three columns display, from left to right: the root-mean-square velocity ($v_{\rm rms}$) normalized by the upstream total Alfv\'en speed ($v_{A,0}$); the kinetic energy ($E_{\rm kin} = \frac{1}{2}\rho v^2$) normalized by the initial total magnetic energy $E_{\rm mag}(0)$; and the magnetic energy components ($E_{\rm mag} = B^2/8\pi$) normalized by $E_{\rm mag}(0)$. Solid and dashed lines in the left and middle columns denote the ion and neutral fluids, respectively. The red and blue lines denote the low-resolution ($8192\times2048$ cells) and high-resolution ($16384\times4096$ cells). In the right column, the solid lines represent the energy of the reconnected magnetic field component ($B_y$, corresponding to magnetic plasmoids), while the dashed lines track the reconnecting field component ($B_x$). The background shading delineates the linear growth stage (light green) and the subsequent non-linear stage (pale yellow).}
    \label{fig:energy_res}
\end{figure*}

To assess numerical convergence, we repeated the $\xi=10^{-2}$ run at half the fiducial resolution ($8192\times2048$ cells; red) and compared it with the fiducial $16384\times4096$ run (blue) in Fig.~\ref{fig:energy_res}. The two runs agree well through both the linear and nonlinear stages. The lower-resolution case shows slightly larger RMS velocity and kinetic energy during the linear and early nonlinear phases, together with a marginally higher growth rate, consistent with modest resolution-dependent numerical dissipation. The magnetic-energy components are nearly indistinguishable, with both the reconnecting field $B_x$ and plasmoid field $B_y$ overlapping throughout the evolution. The displayed global diagnostics exhibit agreement between the two resolutions. However, we note that the density and current-density peaks have resolution dependence; the low-resolution run shows smaller peaks. 

\begin{figure*}[htbp!]
\centering
\includegraphics[width=0.99\linewidth]{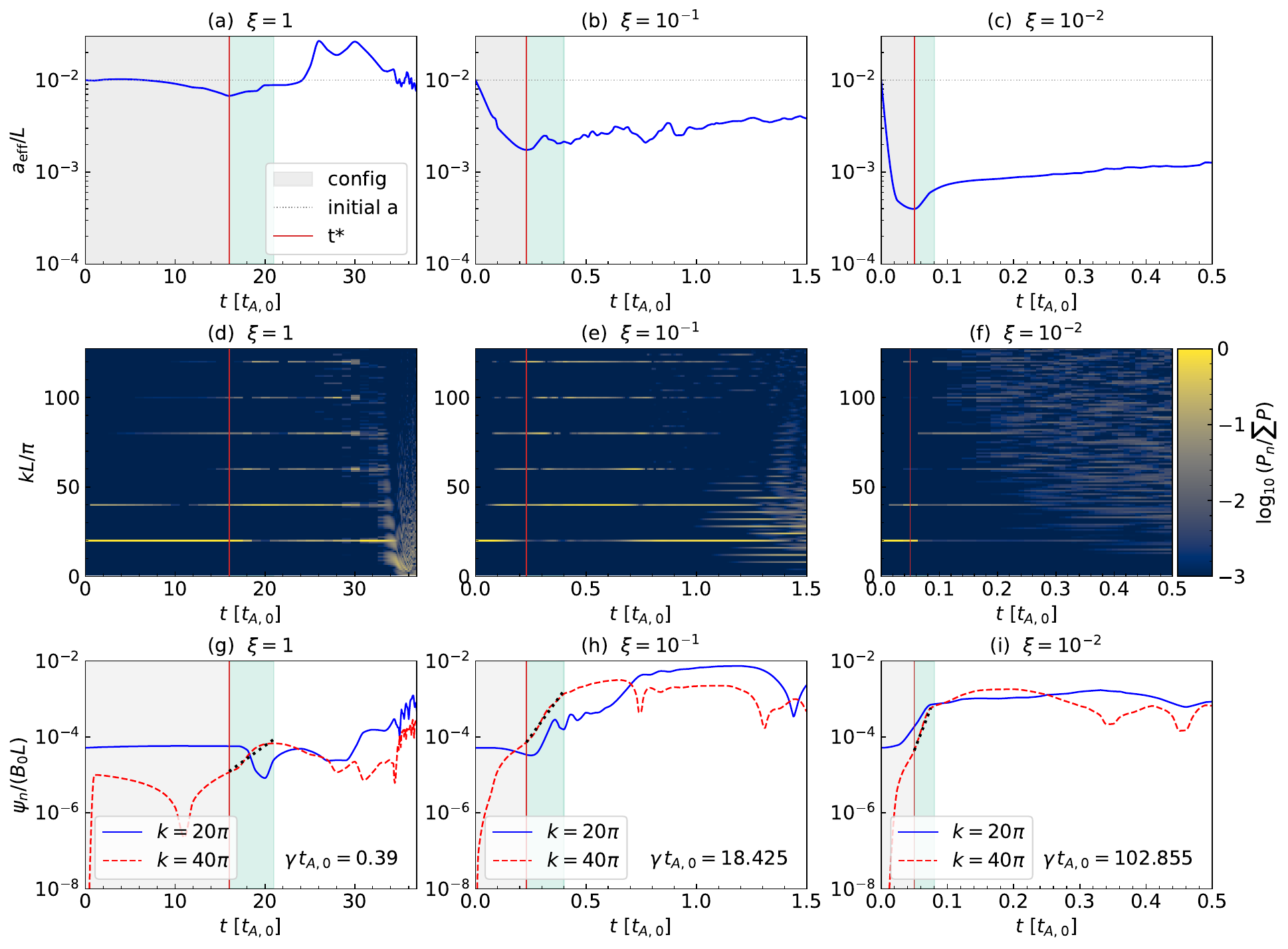}
\caption{Linear-tearing diagnostics for $\xi=1$ (left), $10^{-1}$ (middle), and $10^{-2}$ (right). Top: effective current-sheet half-width $a_{\rm eff}$; the horizontal dotted line marks the initial width, the red vertical line marks the configuration time $t^\ast$, and the shaded region denotes the adopted modal-growth interval. Middle: normalized Fourier power of the normal magnetic-field component $B_y(x,y_0,t)$ as a function of wavenumber and time. Bottom: normalized modal amplitudes for the initially seeded $kL=20\pi$ mode (solid) and the unseeded $kL=40\pi$ mode (dashed). The quoted $\gamma t_{A,0}$ values are obtained by fitting the $kL=40\pi$ mode over the shaded intervals.}
\label{fig:tearing}
\end{figure*}

\section{Linear tearing estimates for the configured current sheets}
\label{sec:tearing_estimates}

Ambipolar transport modifies the current sheet before tearing develops. We therefore compare the measured modal growth with tearing theory evaluated from the locally configured sheet, separating the time required for sheet thinning from the growth subsequently supported by that sheet. We define the instantaneous sheet center $y_0(t)$ from the zero crossing of the $x$-averaged reconnecting field, $\overline{B}_x(y,t)$, nearest the reference midplane. The upstream field $B_{\rm up}$ is measured from the two bands
$0.05L\leq |y-y_0|\leq0.08L$. The effective Harris-sheet half-width is
\begin{equation}
a_{\rm eff}(t)=
\frac{B_{\rm up}(t)}
{\left|\partial\overline{B}_x/\partial y\right|_{y_0}},
\label{eq:effective_sheet_width}
\end{equation}
which recovers the half-width of an exact Harris profile. We define the configuration time $t^\ast$ as the first minimum of $a_{\rm eff}$ following significant sheet thinning.

To characterize tearing growth, we Fourier transform $B_y(x,y_0,t)$ along the instantaneous sheet midplane. For mode number $n$,
\begin{equation}
\psi_n(t)=
\frac{2|\widehat{B}_{y,n}(t)|}{k_nB_0L},
\qquad
\gamma_n^{\rm fit}t_{A,0}
=
\frac{d\ln\psi_n}{d(t/t_{A,0})},
\label{eq:modal_growth_fit}
\end{equation}
where $k_n=2\pi n/L_x$. Since $L_x=2L$, $n=40$ corresponds to $kL=40\pi$.

Fig.~\ref{fig:tearing} quantifies the current-sheet configuration and subsequent modal growth. The effective half-width reaches its first minimum at
$t^\ast/t_{A,0}=16.0$, $0.23$, and $0.05$ for
$\xi=1$, $10^{-1}$, and $10^{-2}$, respectively, with
$a_{\rm eff}/L=6.75\times10^{-3}$,
$1.75\times10^{-3}$, and
$3.97\times10^{-4}$.
Thus, relative to the fully ionized case, ambipolar transport produces sheets thinner by factors of $\simeq3.9$ and $17$ for $\xi=10^{-1}$ and $10^{-2}$.

The Fourier spectra in Fig.~\ref{fig:tearing} show that several modes grow simultaneously after sheet configuration. The $kL=20\pi$ mode is imposed as the initial perturbation and is therefore not used to measure spontaneous tearing growth. We instead compare the unseeded $kL=40\pi$ mode, which is clearly identifiable in all three runs. Over the linear tearing regime, in the order $\xi=(1,10^{-1},10^{-2})$, its fitted growth rates are
\begin{equation}
\gamma_{40\pi}^{\rm fit}t_{A,0}=(0.39, 18.4, 102.9)
\end{equation}
, respectively. The measured growth rates are fitted from the interval $t/t_{A,0}\in[16,21]$, $[0.23,0.40]$, and $[0.05,0.08]$. The same fixed mode therefore grows increasingly rapidly as the ambipolar-drift-configured sheet becomes thinner. 

All quantities entering the theoretical estimates are evaluated at $t^\ast$. The ion and neutral densities are averaged over both upstream bands and the full $x$ direction. In the order $\xi=(1,10^{-1},10^{-2})$, the measured values are
\begin{equation}
\frac{B_{\rm up}}{B_0}=(1.000,0.998,0.986),\quad
\frac{\rho_i}{\rho_0}=(1.000,0.105,0.001),\quad
\frac{\rho_n}{\rho_0}=(0,0.965,0.999).
\end{equation}
We evaluate the tearing growth rate using the measured sheet width $a_{\rm eff}$ for the fixed mode $kL=40\pi$. For a Harris sheet,
\begin{equation}
\Delta'a_{\rm eff}= 2\left[(ka_{\rm eff})^{-1}-ka_{\rm eff}\right],
\end{equation}
where $\Delta'$ measures the tearing drive: larger $\Delta'$ corresponds to a more strongly unstable sheet. The small-$\Delta'$ limit is described by the
FKR branch, while sufficiently large $\Delta'$ enters the Coppi branch \citep{1963PhFl....6..459F,1976SvJPP...2..533C,2020JPlPh..86f5301P}. Because ambipolar drift produces much thinner sheets in the partially ionized runs, the same $kL=40\pi$ mode corresponds to progressively smaller $ka_{\rm eff}$ and
larger $\Delta'$. The fully ionized case therefore lies on the FKR branch, whereas the $\xi=10^{-1}$ and $10^{-2}$ runs lie on the Coppi branch.
Including the ion--neutral effective-inertia correction gives
\begin{equation}
\gamma_{40\pi}^{\rm th}t_{A,0}
\simeq (2.0,\;36.6,\;118.1)
\end{equation}
for $\xi=(1,10^{-1},10^{-2})$, respectively.

Here the instantaneous neutral-ion collision frequencies are $\nu_{ni}t_{A,0} = (-, 1.72\times10^3, 1.64\times10^2)$, in the order $\xi=(1,10^{-1},10^{-2})$, while the fully ionized run has no ion--neutral collision frequencies. Combining these collision frequencies with the growth rates gives
\begin{equation}
\begin{aligned}
    \frac{\gamma_{40\pi}^{\rm fit}}{\nu_{ni}} = (-, 1.07\times10^{-2}, 0.628),\\
    \frac{\gamma_{40\pi}^{\rm th}}{\nu_{ni}} = (-, 2.11\times10^{-2}, 0.720).
\end{aligned}
\end{equation}

The $\xi=10^{-1}$ sheet therefore remains strongly coupled, with $\gamma_{40\pi}^{\rm fit}\ll\nu_{ni}\ll\nu_{in}$. For $\xi=10^{-2}$, the $\gamma_{40\pi}^{\rm fit}$ becomes comparable to $\nu_{ni}$, placing this run near the transition from the strongly coupled to the intermediate regime. Although the theoretical rates agree only approximately with the fitted values, both theory and simulation show a strong increase in growth rate as the configured sheet becomes thinner. This trend is consistent with ambipolar-drift-driven thinning being the primary origin of the enhanced tearing growth in the weakly ionized runs.



\newpage
\bibliography{sample631}{}
\bibliographystyle{aasjournal}



\end{document}